\documentclass[journal]{IEEEtran}
\ifCLASSINFOpdf
\else
\fi
\usepackage[cmex10]{amsmath}
\usepackage{mathtools}
\usepackage[linesnumbered,ruled,vlined]{algorithm2e}
\usepackage{amsmath}
\usepackage[linesnumbered,ruled,vlined]{algorithm2e}
\usepackage{hhline}
\usepackage{amssymb}
\usepackage{multirow}
\newcommand{\R}{\mathbb{R}}
\usepackage{mathtools}
\usepackage{hyperref}
\newcommand*{\doi}[1]{\href{http://dx.doi.org/#1}{doi: #1}}
\usepackage{graphicx}
\usepackage{xcolor}
\usepackage{verbatim}
\newcommand*{\matminus}{%
  \leavevmode
  \hphantom{0}%
  \llap{%
    \settowidth{\dimen0 }{$0$}%
    \resizebox{1\dimen0 }{\height}{$-$}%
  }%
}
\newcommand{\bea}{\begin{equation}}
\newcommand{\eea}{\end{equation}}

\begin{document}
%
\title{ALPS: A Unified Framework for Modeling Time Series of Land Ice Changes}
%
%
%

\author{Prashant~Shekhar, 
        Beata~Csatho, Tony~Schenk, Carolyn~Roberts ~and Abani~Patra~
\thanks{P. Shekhar and A. Patra are with Data Intensive Studies Center, Tufts University, Medford,
MA, 02155 USA. E-mail: \{prashant.shekhar,abani.patra\}@tufts.edu}
\thanks{B. Csatho, T. Schenk and C. Roberts are with the Department of Geology, University at Buffalo, Buffalo, NY, 14260 USA. E-mail: \{bcsatho,afschenk,carolynr\}@buffalo.edu}
}

%
%

\markboth{Journal of \LaTeX\ Class Files,~Vol.~13, No.~9, September~2014}%
{Shell \MakeLowercase{\textit{et al.}}: Bare Demo of IEEEtran.cls for Journals}
%



\maketitle

\begin{abstract}
Modeling time series is a  research focus in cryospheric sciences because of the complexity and multiscale nature of events of interest. Highly non-uniform sampling of measurements from different sensors with different levels of accuracy, as is typical for measurements of ice sheet elevations, makes the problem even more challenging.  In this paper, we propose a spline-based approximation framework (ALPS - Approximation by Localized Penalized Splines) for modeling time series of land ice changes. The localized support of the B-spline basis functions enable robustness to non-uniform sampling, a considerable improvement over other global and piecewise local models. With features like, discrete-coordinate-difference-based penalization and two-level outlier detection, ALPS further guarantees the stability and quality of approximations. ALPS incorporates rigorous model uncertainty estimates with all approximations. As demonstrated by examples, ALPS performs well for a variety of data sets, including time series of ice sheet thickness, elevation, velocity, and terminus locations. The robust estimation of time series and their derivatives facilitates new applications, such as the reconstruction of high-resolution elevation change records by fusing  sparsely sampled time series of ice sheet thickness changes with modeled firn thickness changes, and the analysis of the relationship between different outlet glacier observations to gain new insight into processes and forcing.
\end{abstract}

%
\IEEEpeerreviewmaketitle

%
%
%
%

%

\section{Introduction}\label{Intro}
\IEEEPARstart{T}{he} response of the cryosphere to increasing global temperatures has severe consequences for society. Predictions of the rate of sea-level rise through the next century rely upon accurate understanding and modeling of glacier and ice sheet behavior (e.g., \cite{pattyn:2018hr}). Remote sensing  provides a detailed  record of the Greenland and Antarctic ice sheets, including observations of ice sheet elevation, velocity, and extent.  These observations  revealed dramatic changes of many ice streams and outlet glaciers  since the late 1990s, causing alarming mass loss  (\cite{enderlin:2014jr}, \cite{vanDenBroeke:2016ch}, \cite{theIMBIEteam:2018caa}, \cite{theIMBIEteam:2019}). However, the significant spatiotemporal variability of ice velocity and elevation changes compounded by the irregular distribution of the observations  makes the interpretation of the remote sensing record difficult. While regional trends exist, the different patterns of change within a single drainage basin and between neighboring glaciers indicate that the response of individual outlet glaciers to external forcings is highly modulated by local conditions \cite{csatho:2014jja}. 

A significant mathematical challenge is posed in dealing with irregularly distributed space-time data of highly variable quality that must be modeled for characterizing ice flow, surface properties and behavior. Data acquired by different sensors and missions often have different  coverage and sampling. Data quality can also vary, and data gaps can occur due to system limitations, low atmospheric transmittance, or unfavorable surface conditions. The large size of the ice sheets (e.g., 14 million km$^2$ of Antarctica) prohibits a simultaneous or near-simultaneous observation of an entire ice sheet with high resolution and accuracy. For example, it takes 91 days for NASA's new laser altimetry mission, Ice, Cloud and land Elevation Satellite-2 (ICESat-2), to complete a full cycle of observations  \cite{markus:2017dd}. Therefore, even data from the same mission should be interpolated to obtain snapshots of regional or ice sheet scale conditions, such as Digital Elevation Models (DEM) or annual thickness change rates.

Accurate  spatiotemporal modeling of ice sheet observations serves many purposes, including the monitoring of ice sheet elevation change and  mass loss (e.g.,  \cite{csatho:2014jja}, \cite{schenk2012new}), as well as providing input for data assimilation into numerical ice sheet models \cite{Larour:2014cg}. Current approaches for analyzing time series of ice sheet changes (e.g., elevation, velocity) rely on simple models, such as low-order polynomials (e.g., \cite{schenk2012new}, \cite{Flament:2012bx}, \cite{schroder:18}) or a combination of linear interannual changes and seasonal signals estimated with a trigonometric function (e.g., \cite{Nilsson:2016}). As longer and more detailed time series become available, a need arises to accurately model complex and rapid spatiotemporal changes  and to derive robust error  estimates, which are especially important in sparsely sampled locations. 

Our new approach, ALPS (Approximation by Localized Penalized Splines), is a penalized spline based framework for modeling cryospheric (ice sheets, glaciers, permafrost, sea ice) changes (python code is available online \footnote{\url{https://github.com/pshekhar-tufts/ALPS.git}}). The main objective of ALPS is to model  ice sheet thickness- and elevation-change time series derived from laser altimetry data and DEMs using the  Surface Elevation Reconstruction And Change detection (SERAC)  approach \cite{schenk2012new}.  Using a least-squares approach for determining surface elevations, SERAC  reduces random errors and facilitates the detection of outliers. Therefore, typical  laser altimetry time series are characterized by small elevation errors and few outliers but have irregular temporal sampling, often with large gaps, due
to the varying length and repeat periods of different satellite and airborne missions (e.g., \cite{Schenk:2014}, \cite{Babonis:2016}). 

In this paper, we introduce the working principles of ALPS. After a brief introduction, Section II presents a comparison of global and local fitting approaches to motivate this research, followed by a description of the data used in this study.    We describe ALPS in Section IV where dynamic ice thickness changes in different regions of the Greenland Ice Sheet are used to illustrate the performance of the algorithm. The paper concludes with  two application examples. In
Section V.A
we show how to employ ALPS for  generating a high-temporal resolution (10 days) reconstruction of ice sheet surface changes by combining sparsely sampled altimetry data with a climate model output.
Finally, in Section V.B we use ALPS to investigate the mechanisms controlling changes of the Helheim Glacier in East Greenland between 2001-2010 by analyzing ice thickness, velocity, and calving-front change time series and their derivatives. The applications also include a comparison of ALPS with traditional polynomial models, providing further evidence regarding its capabilities for modeling high resolution, rapid, local and global changes in a stable and robust manner.

\section{ Motivation}\label{motivation}

The currently available  laser altimetry record of more than 25 years of ice sheet elevation changes (1993-2020) enables the monitoring  of  Greenland and Antarctic ice sheets changes on scales ranging from individual outlet glaciers to the entire ice sheet with unprecedented accuracy and detail (\cite{csatho:2014jja, Babonis:2016}). However, as we show here, the widely used simple interpolation and approximation methods are not suitable to exploit the full temporal resolution of the laser altimetry data set, especially in rapidly changing regions with complicated elevation change patterns.

The distribution of laser points obtained from airbone and spaceborne laser altimetry systems is very irregular in both the spatial and temporal domains. Spatially, the points are confined to lines or narrow bands (ground tracks or swaths) with an along-track sampling distance determined by the frequency of the laser's firing rate. In order to derive surface elevation changes, laser altimetry measurements  have to be repeated periodically. Such is the case with NASA's ICESat and ICESat-2 missions, with repeat observations  collected every 91 days along repeat ground tracks in the polar regions (\cite{Zwally:2002vp, markus:2017dd}). Environmental conditions, such as fog, cloud coverage or blowing snow may prevent the laser beam to reach the ice surface, or may render a wrong position for the point from where the laser was reflected (blunder). The spatiotemporal pattern of repeat satellite groundtracks and airborne laser altimetry swaths is often quite complicated   (e.g., Fig. 1 in \cite{Schenk:2014}) resulting in significantly different temporal sampling in  neighboring locations.

Time series calculated from repeat laser altimetry are useful in many applications, including monitoring ice sheet elevation change and mass loss, or they may serve as input to ice sheet models. Considering the irregular spatiotemporal distribution of laser points, amplified by including laser altimetry data from different sensors, as well as DEMs, the interpolation of the diverse data sets requires the application of a sophisticated approach. 
Current methods use  linear combinations of simple basis functions having global support, including   linear  (\cite{pritchard2012antarctic, shepherd2010recent}), quadratic  \cite{wingham2009spatial} and variable degree    (\cite{schenk2012new, paolo2016constructing}) polynomial models. To model both the elevation change trend and its seasonal variation,  \cite{zwally2005mass} and \cite{slobbe2008estimation} used a combination of a linear trend and a trigonometric function.

\begin{figure}[h]
\centering
\includegraphics[width=3.3in]{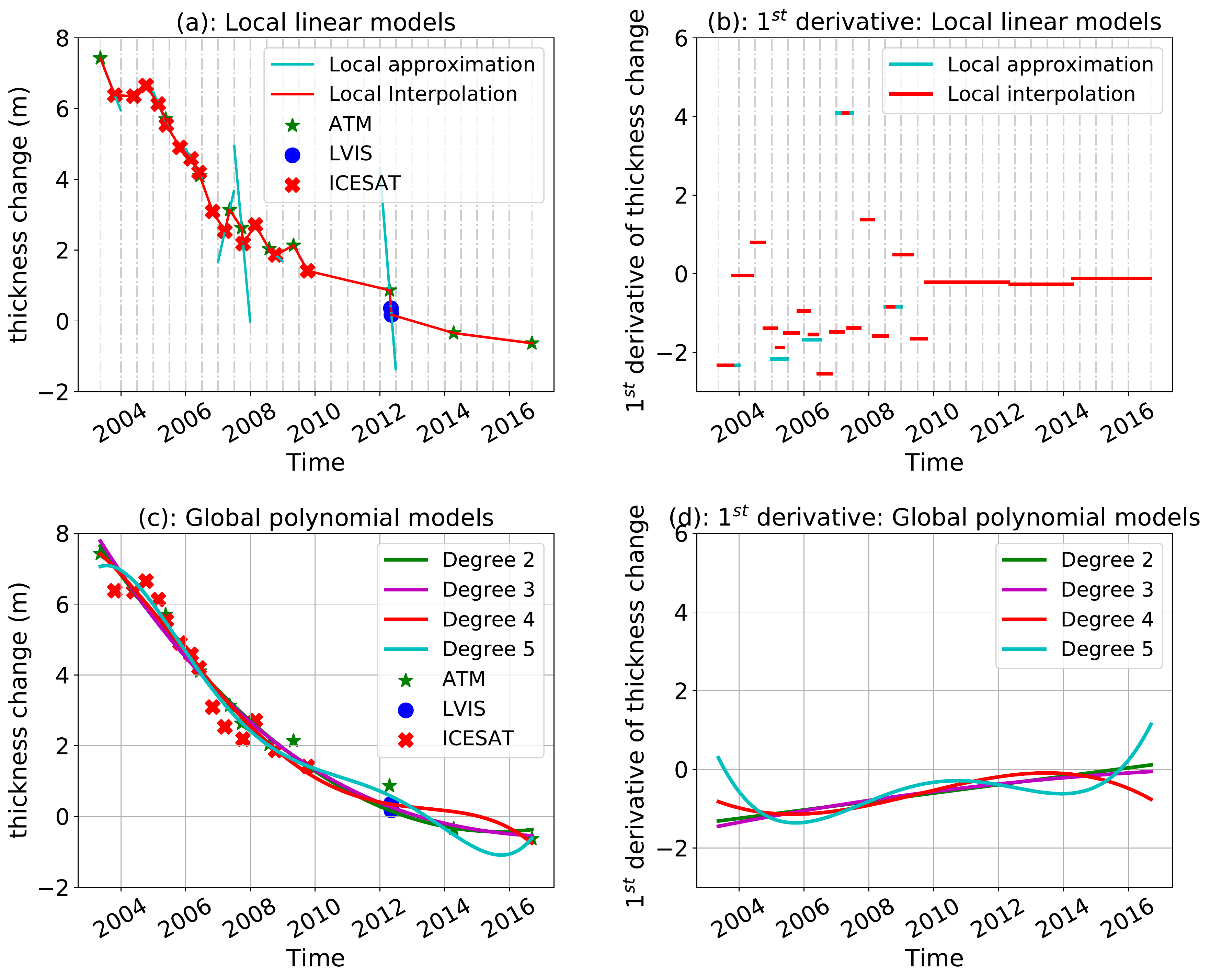}
\caption{Comparision of methods using polynomials for modeling   thickness change time series derived from airborne (ATM, LVIS) and spaceborne (ICESat) altimetry observations, TS:0. (a and b):  Local linear models  (a)  local linear approximation  with half-annual resolution (light blue) and local linear interpolation  (red), (b) corresponding first derivatives. (c and d): Global polynomial approximation models (degree 2,3,4 and 5) (c) predictions, (d)   corresponding first derivatives.}
\label{comparison1}
\end{figure}

In Fig.~\ref{comparison1} we visually demonstrate the shortcomings of these methods using one of  the more than 100,000 time series we have calculated in Greenland \cite{Babonis:2016}. This time series, labeled as TS:0 (Fig.~\ref{Locmap} and Table~\ref{timeseries}), depicts the  change in ice sheet thickness  at  ~1700 m elevation (on WGS-84 ellipsoid) in the drainage basin of the Helheim Glacier.  It clearly shows the irregular temporal sampling characteristic of a multi-sensor time series: fairly dense during the ICESat satellite altimetry mission (2003 to 2009, \cite{Zwally:2002vp}) and  much longer sampling intervals during the Operation IceBridge (OIB)  mission  that employed the Airborne Topographic Mapper (ATM, \cite{Krabill:2002wg}) and Land, Vegetation, and Ice Sensor (LVIS, \cite{Hofton:2008ff}) airborne laser altimetry systems (2009-2019). Evident is the relatively large error of ICESat observations (Fig.~\ref{comparison1}, green crosses), particularly noticeable from 2007 to 2009.

Fig.~\ref{comparison1}(a) compares the performance of a local linear interpolation (red lines connecting observations) and a half-annual resolution, local linear approximation model  (blue line, intervals span January 1-June 30 and July 1-December 31 of each year). In case of our noisy data the linear interpolation  fits closely to the noise  that is especially evident in the very noisy derivatives (Fig.~\ref{comparison1}(b)).  Imposing a half-year resolution for a linear approximation is also problematic because of the lack of sufficient support to calculate the fitted lines and thus provide estimates in most half-year periods. Choosing the window size dynamically would solve the problem of having windows without data, but still produces discontinuities, making the computation of analytical derivatives impossible. 
 Also, it brings an additional challenge of deciding the window size, i.e., determining the "break times", separating the periods of linear behavior, automatically, which is a difficult problem  with no obvious best choice \cite{jamali2015detecting}.

Fig.~\ref{comparison1} (c) depicts the same time series, approximated by polynomials of varying degrees. We recognize some well-known problems with polynomials, such as the oversmoothing by low order polynomials (period 2003 to 2008) rendering them unsuitable  for  modeling short-term, rapid variations, or multiple periods of subtle, but significant ice thickness changes, behaviors often observed on outlet glaciers  (Table~\ref{timeseries}, TS:1-3, 6a).  Also, higher-order polynomials exhibit a sensitivity  to observation errors resulting in an oscillatory behavior, especially in periods with low sampling rate, such as the epoch between 2012 and 2016. This is  evident in Fig.~\ref{comparison1} (d) showing rapid changes in the  first derivative during time periods of relatively little change (Fig.~\ref{comparison1} (c)). Yet another drawback of global approximation can be seen in the propagation of local perturbations through the entire temporal domain.

Another factor we have to take into account is the number of time series we are dealing with for certain applications, such as mass balance studies for the entire ice sheet of Greenland with over 100,000 thousand time series repeatedly calculated   when new data becomes available (\cite{csatho:2014jja, Babonis:2016}). Keeping this in mind we need robust and autonomous algorithms as it is impossible for human operators to go through these massive data sets.

In summary we conclude that the shortcomings of commonly used methods  pose severe limitations for 
faithfully transforming our discrete time series into suitable analytical forms. 
We need a robust, autonomous method capable to fit locally to the data and preventing 
the spread of local disturbances into the entire time domain. Such is the case 
with ALPS that uses localized B-spline bases, preserves smoothness and 
differentiability and copes well the with the typical irregular sampling
of our altimetry time series. ALPS provides a unified framework for converting discrete
time series into analytical forms, without the need of subjective
selections, even in cases of noisy, non uniformly sampled data 
contaminated by blunders.

\section{Data Sets and Preprocessing}\label{Data}

Fig.~\ref{Locmap} and Table~\ref{timeseries} summarize the information about all ice sheet elevation, velocity and terminus location time series used in this study.

\subsection{Laser Altimetry and DEMs}

Several different remote sensing techniques, including repeat photogrammetry, laser or radar altimetry and interferometric radar are used to determine ice sheet surface elevation changes \cite{Cuffey:2010}. Airborne and spaceborne laser altimetry measurements are frequently used because of their high accuracy and spatial resolution. 

NASA's  ATM and LVIS airborne laser altimetry systems (\cite{Krabill:2002wg, Hofton:2008ff})  collected  data along repeat transects between 1993-2019 as part NASA's Program for Arctic Regional Climate Assessment (PARCA, 1993-2008) and OIB (2009-2019) missions. Combined with the repeat, synoptic coverage  of the  ICESat (2003-2009, \cite{Zwally:2002vp}) and ICESat-2 missions (2018-present, \cite{markus:2017dd}) these observations enabled the monitoring of Greenland and Antarctic ice sheet changes on scales ranging from individual outlet glaciers to  entire ice sheets with unprecedented accuracy (\cite{csatho:2014jja, Babonis:2016, Smith:2020en}.

For this study, we used ATM (\cite{Studinger:2014}) and LVIS \cite{Blair:2019} airborne and ICESat (\cite{Zwally:2014}) spaceborne laser altimetry data, all obtained from the National Snow and Ice Data Center. Time series of ice sheet elevation changes were derived from the original laser altimetry point measurements using  the  Surface Elevation Reconstruction And Change detection (SERAC)  approach \cite{schenk2012new}.

SERAC is an area-based method, developed to calculate surface changes from altimetry measurements that are not repeated at the  same  locations. Instead  of collocating the point observations to the same spatial location before calculating changes (e.g., \cite{Zwally:2011eg}), SERAC determines the elevation change time series by simultaneously reconstructing the shape and elevation change  from  all  observations (spanning the entire observation period)  within a spatial neighborhood  (\cite{schenk2012new}). Locations of the surface patches are determined automatically from the spatial distribution of the data. A surface patch size of 1 by 1 km was used to generate the time series presented in this study \cite{Babonis:2016}. The solution assumes that the shape of the surface patch does not change, only its elevation.

The large redundancy of the observations enables the formulation of the problem as a least-squares adjustment, providing rigorous error estimations. Typical formal errors of the time series in the  accumulation zone of the ice sheet at higher elevations are about $\pm0.02$ m, similar to the error of individual laser observations under ideal conditions \cite{Fricker:2005jba}. At lower elevations, the error increases and reaches values of $\pm 1.0$ m or even larger, because of increased slope and roughness (e.g., crevasses). 

In our typical altimetry processing workflow, we perform temporal interpolation after the removal of ice thickness changes due to surface processes   \cite{csatho:2014jja, Babonis:2016} to mitigate the issue of sparse temporal sampling \cite{csatho:2014jja, Babonis:2016}. Ice sheet elevation changes include seasonal variations and other short-term changes that are often poorly sampled, in particular during periods without satellite laser altimetry missions. However, thickness changes due to changing ice velocity (ice dynamics), computed by removing the changes due to surface processes,  typically vary on an interannual or decadal scale, and seasonal variations are often negligible. Thus, robust modeling of dynamic thickness changes can be achieved despite the uneven temporal sampling. Note that all dynamic thickness change time series are normalized to a reference time, August 31, 2006, selected in the middle of the ICESat mission (Figs.~\ref{comparison1}, \ref{gcv}-\ref{new_2}, \ref{app21}-\ref{app22}).

To create the detailed elevation time series of the main trunk of the Helheim glacier (Section~\ref{helheim}.B) laser altimetry data were combined with DEMs. The systematic errors of the DEMs were removed using altimetry time series on the glacier derived by the SERAC method as control information  \cite{Schenk:2014}. 

A dozen DEMs generated were generated from ASTER (Advanced Spaceborne Thermal Emission and Reflection Radiometer) satellite imagery using the  MicMac ASTER (MMASTER) DEM correction 
 approach \cite{LucGirod:2017eb}. The MMASTER method improves the matching between ASTER bands 3N and 3B by estimating the cross-track direction parallax error and removing most cross-track jitter before the DEM generation, resulting in an improvement of tens of meters of the vertical accuracy of the ASTER DEMs. The MMASTER DEMs were generated from the raw L1A ASTER scenes by C. Nuth (pers. comm., 2017), and aimed at capturing the elevation change on Helheim Glacier at times when laser altimetry was sparse/missing (e.g., 2004).

One Satellite Pour l’Observation de la Terre  5 (SPOT-5) DEM, acquired in 2007, was also used. It was made available via the SPOT-5 stereoscopic survey of Polar Ice: Reference Images and Topographies (SPIRIT) program, which provided orthophotos and DEMs of ice sheet coastal regions as a contribution to the fourth International Polar Year \cite{Korona:2009bw}. The data set  was downloaded from the Theia website (theia-landsat.cnes.fr).


\subsection{Ice velocity}
The velocity time series used for characterizing the recent mass loss and thinning of the Helheim Glacier  are  from two freely available ice-velocity datasets. The optically derived dataset from the Technical University of Dresden (TUD; \cite{Rosenau:2015bg}) utilizes the Landsat record (1999-2014), and thus, provides velocity data at monthly, sometimes sub-monthly intervals during a critical time period for Helheim Glacier. The formal error of the TUD velocity data set is 136 m/yr (RMS). The other velocity data set is derived from 
Synthetic Aperture Radar measurements acquired by the twin satellites of the German Aerospace Center's  TerraSAR-X (TSX) mission \cite{Joughin:2019}. This data set has very high spatiotemporal resolution, existing at 11- to 33- day intervals over a large area of the glacier at 100 m grid posting with very low errors (RMS error: 11 m/yr).  TUD data were obtained from the TUD Geodetic Data Portal, and TSX data were downloaded from the National Snow and Ice Data Center.

\subsection{Terminus position}
To characterize changes in the location of the terminus of the Helheim Glacier, we used a high temporal-resolution normalized terminus position time series   derived from 2001-2010 Moderate Resolution Imaging Spectroradiometer (MODIS) satellite imagery from \cite{Schild:2013ie}.

\begin{figure}[h]
\centering
\includegraphics[width=3.4in]{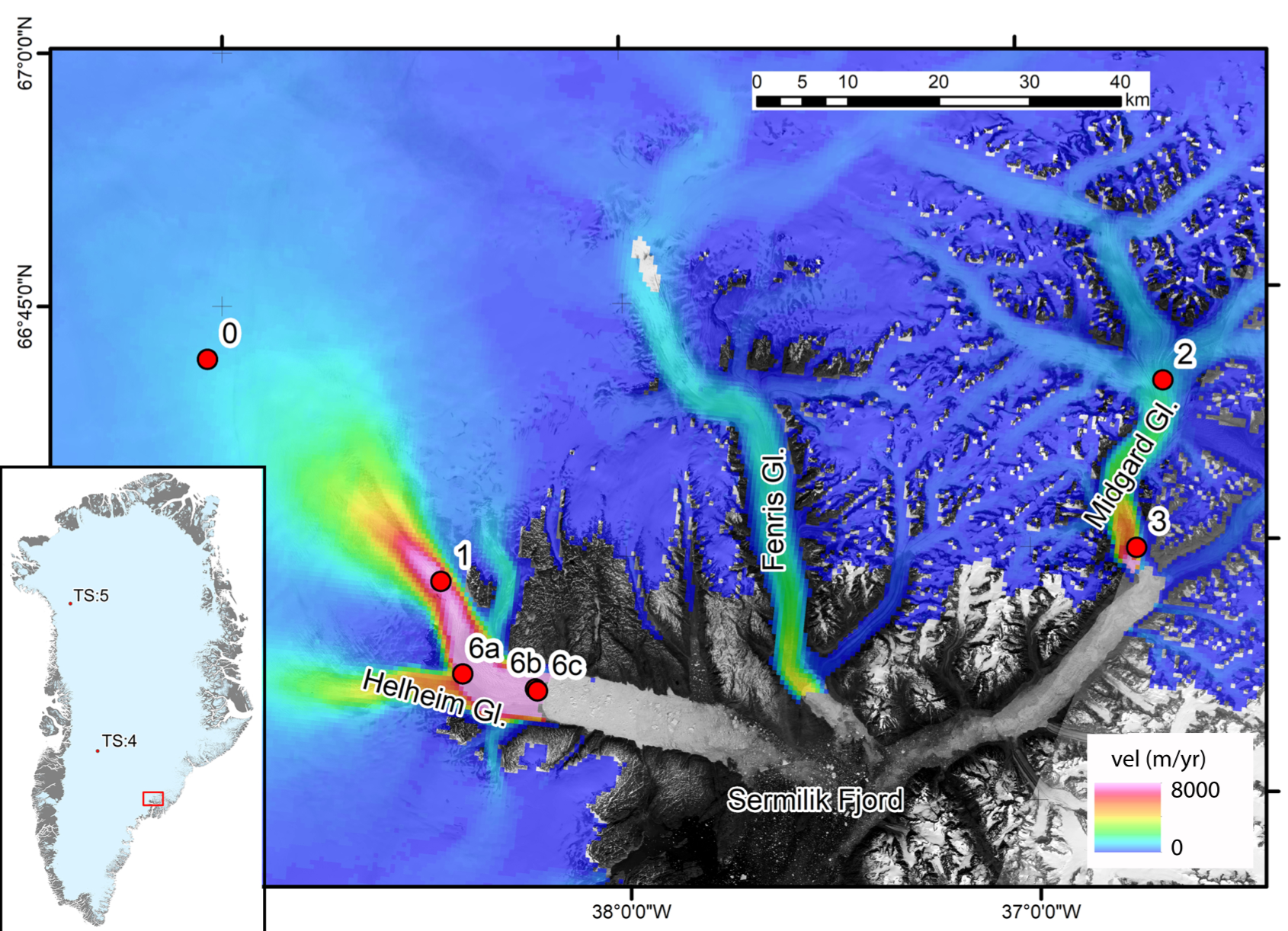}
\caption{Ice velocity map  of Helheim, Fenris and Midgard glaciers in east Greenland with  locations of time seres 0-3, and 6a-c.  The region is marked by a red rectangle in the inset map of the Greenland Ice Sheet, also showing TS:4-5 in western Greenland. 1995-2015 ice velocity mosaic is from \cite{Joughin:2016} with a Landsat-8 imagery dated June 16, 2018 in the background.}
\label{Locmap}
\end{figure}

\begin{table*}[h]
    
    \centering
    \small
  \setlength{\tabcolsep}{3pt} 
    \begin{tabular}{c|c|c|r|r|r|l|l|l|l|l}
        TS  & Lat & Lon & H$_{start}$  & $\Delta$H  & Duration  & Sensors & Description & Data & Sections & Figures  \\
         & (deg) & (deg) &  &  &     & Sources  &  &     \\
        \hline
        \hline
              &  &   &   &   &  &  &  &  &   \\
         0 & 66.698 &  -39.036 &	1745.2	& 	19.7 &	2003-2016 &	ICESat, ATM, LVIS &	 Helheim Gl. upstream,   & \cite{Zwally:2014, Studinger:2014} & \ref{motivation}, \ref{comp_method}, & 1, 9 \\
         &  &  &  &  &  &  &subglacial lake drainage  & \cite{Blair:2019,Roberts:2019} &  & \\
            \hline
              &  &  &   &   &     &  &  &  &   \\
              
         1 & 66.4779 &	-38.4590 &	755.3	& 	68.8 &	1998-2017 &	ICESat, ATM, LVIS &	 Helheim Gl.   & \cite{Zwally:2014, Studinger:2014} & \ref{fittingpspline}, \ref{implementation} & 4, 5, 7(a) \\
         &  &  &  &  &  &  &main branch  & \cite{Blair:2019} &    
          \ref{comp_method}  & 10 \\
            \hline
              &  &  &   &   &     &  &  &  &   \\
         2 & 66.6603 &  -36.6566 &	1028.1 & 94.0	& 2003-2017	& ICESat, ATM	& Midgard Gl. & \cite{Zwally:2014, Studinger:2014} & \ref{fittingpspline} &  6(a)  \\
            &  &  &   &   &    &  & mid glacier &  &   \\
               \hline
              &  &  &   &   &     &  &  &  &   \\
        3 & 66.4961 &  -36.737 &	528.6 & 228.7	& 2004-2017	& ICESat, ATM, LVIS	& Midgard Gl. & \cite{Zwally:2014, Studinger:2014} & \ref{fittingpspline} & 6(b), 7(b) \\
            &  &  &   &   &   &    & near terminus & \cite{Blair:2019} &   &  \\
             \hline
              &  &  &   &   &   &    &  &  &   \\
         4 & 68.9665 &  -45.7304 &	2035.3 & 3.8	& 2003-2017	& ICESat, ATM	& Jakobshavn Gl. & \cite{Zwally:2014, Studinger:2014} & \ref{out} & 8 \\
             &  &  &   &   &   &    & slow ice &  &   \\
             \hline
              &  &  &   &   &   &    &  &  &   \\
      5 & 76.3150 &  -55.3177 &	1922.5 & 0.3	& 1994-2015	& ICESat, ATM	& Nansen Gl., upstream & \cite{Zwally:2014, Studinger:2014} & \ref{firn} & 11\\
          &  &  &   &   &     &  & slow ice &  &   \\
         \hline
           &  &  &   &   &   &    &  &  &   \\
        6a & 66.3861 &  -38.4073 &	470.2 & 94.9	& 2001-2010	& ICESat, ATM, 	& Helheim Gl. & \cite{Zwally:2014, Studinger:2014}, & \ref{helheim} & 12(a), 13\\
             &  &  &   &   &     & ASTER, SPOT &  confluence & \cite{LucGirod:2017eb, Korona:2009bw} &   \\
              &  &  &   &   &     &  & medial moraine &  &   \\
             \hline
                &  &   &   &   &  &  &  &  &   \\
         6b & 66.3706 &  -38.2284 &	260.5 & 151.6	& 2001-2010 & Landsat (TUD)	& Helheim Gl.  & \cite{Rosenau:2015bg} & \ref{helheim} & 12(b)\\
         
             &  &  &   &   &     & InSAR (TSX) & trunk  & \cite{Joughin:2019}  &   \\
                 \hline
                &  &   &   &   &  &  &  &  &   \\
        6c & 66.3685 &  -38.2232 &	 & 	& 2001-2010	& MODIS	& Helheim Gl.   & \cite{Schild:2013ie} & \ref{helheim} & 12(c), 13\\
        &  &  &   &   &     &  & calving front max. retreat &   \\
           &  &  &   &   &     &  &  &   \\
    \end{tabular}
    \caption{Time series (TS) of surface elevation (0-6a), velocity (6b), and calving front location (6c) used in this study. H$_{start}$ represents the surface elevation at the beginning of time series, and $\Delta$H shows its change during the span of the  time series. Surface elevations are given in $meters$ on the WGS-84 reference ellipsoid.  See Fig.~\ref{Locmap} for locations of the time series.}
    \label{timeseries}

\end{table*}



\section{Methodology}\label{methodology}

In this section we describe our proposed ALPS framework that builds on Penalized Spline (P-spline) \cite{eilers1996flexible} approximations for time series modeling. We use B-spline basis functions to capture local information in the data, with an added penalty on the separation of coefficients to provide adequate smoothness and robustness to noise.

\subsection{B-spline approximation}\label{bspline}

For the formulation of B-splines we follow \cite{de1978practical} and \cite{piegl2012nurbs}. We define  the {\it knot }vector, $U = \{u_0, u_1,...u_m\}$ as  a non-decreasing sequence of real numbers such that $u_a \leq u_{a+1}$ for $a = 0,1..,m-1$. The coefficients $u_a \in U$ are known as knots and determine the length of individual sections on the curve. Let $p$ be the degree of the i-th B-spline basis we want to create. 
For $u_0 \leq t < u_m$ we have

\begin{equation}
\begin{multlined}
B_{i}(t;p) =\\
 \frac{(t-u_i)B_{i}(t;p-1)}{u_{i+p} - u_i} + \frac{(u_{i+p+1} - t)B_{i+1}(t;p-1)}{u_{i+p+1}-u_{i+1}}
\label{basis}
\end{multlined}
\end{equation}
\begin{eqnarray}
B_{i}(t;0) = \begin{cases} 
	1 & u_i \leq t < u_{i+1} \nonumber \\
	0 & otherwise
\end{cases}
\end{eqnarray}

As a consequence of the definition of
$B_i(t;0)$,  $B_i(t;p)$ is  non-zero only for $t \in [u_i,u_{i+p+1})$. 
As $t$ varies from $u_0$ to $u_m$, different basis functions become non-zero at different values of $t$ governed by (\ref{basis}).
Let the domain under consideration $[x_{min},x_{max}]$  be divided into $m$ intervals by $m + 1$ knots with $u_0 = x_{min}$ and $u_{m} = x_{max}$. From (\ref{basis}) it follows that for a B-spline basis of degree $p$ we need  $m + 2p + 1$ knots and the number of basis functions is $c = m + p$, according to the definition given in \cite{eilers1996flexible}.  More detailed information about B-spline basis functions can be found in \cite{de1978practical} and \cite{piegl2012nurbs}.

\begin{figure}[h]
\centering
\includegraphics[width=3.3in]{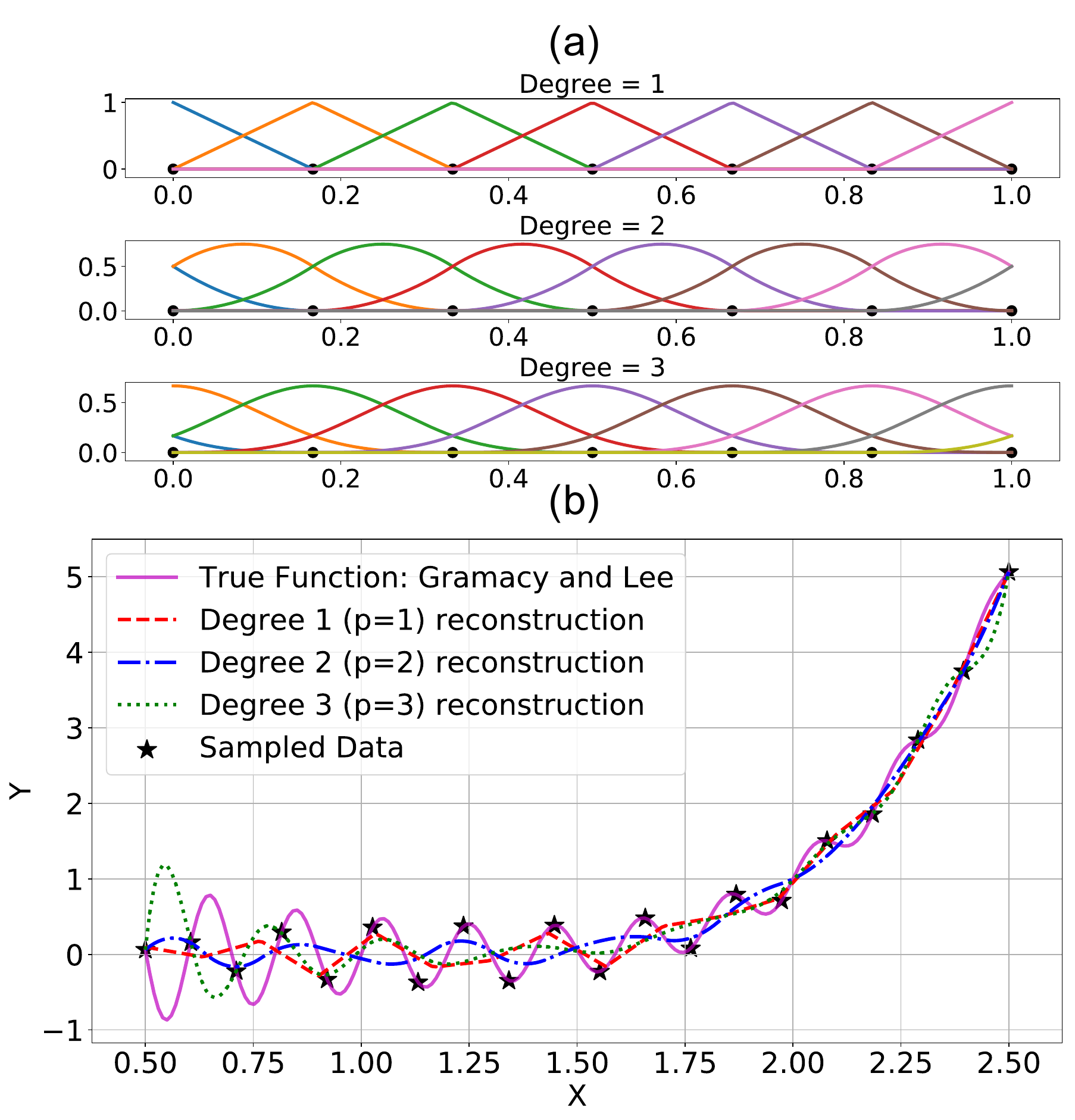}
\caption{(a) B-spline basis functions with 6 (m = 6) sections and degree (p) = 1, 2 and 3 shown with different colors to emphasize their local support. Here the support and height of individual basis function comes from (\ref{basis}) with x axis scaled as per the duration of a time series and y axis showing the value (height) of the basis functions in B.  (b)  B-spline approximations of Gramacy and Lee 2012 test function \cite{gramacy2012cases}.  Black stars mark the samples of the test functions, and reconstructions using B-spline basis functions with degree 1, 2 and 3 are shown by red dashed, blue dashed and green dotted  lines, respectively.}
\label{4}
\end{figure}

We illustrate basis functions  with degrees 1 to 3 and with uniformly spaced knots in Fig.~\ref{4}(a). Here $u_0$ = 0 and $u_m$ = 1 with m = 6. Based on the recursive formula (\ref{basis}), at p = 1, we get a straight line basis with local support. However with increasing degree, the basis functions become nonlinear functions of parameter t. The solid circles on the horizontal axis show the distribution of the knots. Note that there are $m+2p+1$ knots with p knots before $u_0$ and p knots after $u_m$ (not shown in the figure). As demonstrated by the figure, with increasing degree, the basis functions become smoother with a higher capability for modeling non-linear behavior (due to increased degrees of freedom). However, the increased flexibility can also lead to overfitting (which is handled by ALPS using a separate penalty operator).

For modeling directly using the B-spline bases, consider a function $f(t)=\sum_{i=0}^{c-1} B_{i}(t;p)\theta_{i}=B(t) \Theta,$ where $B(t) = \{B_{i}(t;p), 0 \leq i \leq c-1 \}$ are a set of B-spline basis as defined in  (\ref{basis}). In this equation $t$ is the regressor and $\Theta=\{ \theta_{i}\}_{i=0,..,c-1}$ are the unknown coefficients. If $y$ denotes the response variable, then the resultant model becomes

\begin{align}\label{modelgcv}
y(t)& =f(t) + \epsilon; \quad \epsilon \sim N(0,\sigma^2 I)
\end{align} 
For the ice sheet thickness change models, $f(t)$ (modeled as $B(t) \Theta$) in (\ref{modelgcv}) is the approximation which maps the time of the observations $t$ to the corresponding ice thickness change. The error model $\epsilon$ has a mean of zero and a variance $\sigma^2$ ($I$ is the identity matrix with dimensions equal to the number of samples). The least squares formulation for estimating $ \Theta$, given a vector of data ${\bf y}=\{y_j\}_{n \times 1}$  at   observation times $t_j$ and $B=[b_{ji}]$ (where B is a $n \times c$ matrix and $b_{ji}=B_i(t_j;p) $), is

\begin{eqnarray}\label{leasq}
\min_\Theta S &=& \min_\Theta(y - B \Theta)^T (y - B \Theta) \\
\Rightarrow \hat{\Theta} &=&(B^T B)^{-1} B^Ty
\end{eqnarray}

For illustrating the performance of B-spline approximation with variable degree, we selected the Gramacy and Lee test function  for its global and local patterns, which make it challenging for any approximation procedure to  model it accurately (Fig.~\ref{4}(b), \cite{gramacy2012cases}). As shown in Fig.~\ref{4}(b), the lower degree basis functions do not capture the details of the signal, while the higher degree approximations include undesired fluctuations,  not supported by the data. This limitation of the direct approximation by B-spline basis functions motivates the introduction of penalized splines that incorporate a built-in mechanism to suppress spurious fluctuations occuring at a higher degree.   Moreover, approximations with B-spline bases are  prone to stability issues, especially when undersampling occurs. For example, from a sparse sampling, the degree 3 B-spline approximation  reconstructs the first minimum of the Gramacy and Lee test function as a maximum (Fig.~\ref{4}(b)). In applications such as, modeling of ice sheet thickness  change  from an altimetry data set, this issue of mirroring and shifting peaks in the time series can lead to wrong inferences about thickening or thinning.

\subsection{Penalized Splines}\label{psplines}
Eiler and Marx  \cite{eilers1996flexible} proposed a penalty  on the difference of the adjacent coefficients of the basis functions and minimize the following penalized sum of squares \cite{lee2010smoothing}

\begin{eqnarray}
\min_\Theta S_p &=& \min_\Theta \{(y - B\Theta)^T(y - B\Theta) + \Theta^T P \Theta \}  \label{cost}
\end{eqnarray} 

While it is possible to find an exact minimizer of (\ref{cost}), we still need to obtain an estimate of P (represented as $\hat{P}$). The solution for $\Theta$ (penalized least squares) in (\ref{cost}) is then given as
\begin{eqnarray}
\hat{\Theta} &=& (B^T B + \hat{P})^{-1} B^T y \label{theta}
\end{eqnarray} 

This constrained minimization smoothes the fit and keeps only large, abrupt changes. Following \cite{eilers1996flexible,lee2010smoothing} we apply discrete penalties and  penalize the difference in the coefficients of the adjacent basis functions. Since the number of basis functions is $c$, the penalty matrix has a dimension of $c \times c$ and can be written as 

\begin{equation}
P = \lambda (\Delta^q)^T \Delta^q,
\label{lammb}
\end{equation}
where $\Delta^q$ is the difference operator of order $q$ (with $q$ being the order of penalty) operating on $\Theta$ and $\lambda$ is a  hyperparameter. For a vector of regression coefficients $\Theta$, the difference operator is defined recursively (with increasing penalty order q) as

\[\Delta^1 \theta_i = \theta_i - \theta_{i-1}\]
\[\Delta^2 \theta_i  = \Delta^1(\Delta^1 \theta_i) = \theta_i - 2 \theta_{i-1} + \theta_{i-2}\] 
\[\vdots\]
\[\Delta^q \theta_i = \Delta^1 (\Delta^{q-1} \theta_i ) \]

In matrix form $\Delta^q$ is represented as $D_q$. For example for first and second order difference (q = 1 and 2) and c = 5, the D matrices are

\[D_1 = \begin{bmatrix} \matminus1 &1& 0& 0& 0 \\0 & \matminus1 & 1 & 0 & 0 \\  0 & 0 & \matminus1 & 1 & 0 \\ 0& 0& 0& \matminus1 &1 \end{bmatrix},
D_2 = \begin{bmatrix}1& \matminus2 &1 &0 &0 \\ 0& 1& \matminus2& 1& 0\\ 0& 0 &1 &\matminus2& 1 \end{bmatrix}\]
%

 Illustrating for $c=5$ and $ q=1$, we have $\Theta = [\theta_1,\theta_2,\theta_3, \theta_4, \theta_5]^T$ and the penalty term $\Theta^T P \Theta$ in (\ref{cost}) using 
 $D_1^TD_1$ is given as

\bea \Theta^TP\Theta = \lambda \Big[(\theta_2 - \theta_1)^2 + (\theta_3 - \theta_2)^2 + (\theta_4 - \theta_3)^2 + (\theta_5 - \theta_4)^2  \Big] \label{illustr_1st} \eea

Since, (\ref{illustr_1st}) involves only first order differences, it is called a first order difference penalty.

\begin{figure}
\centering
\includegraphics[width=3.1in]{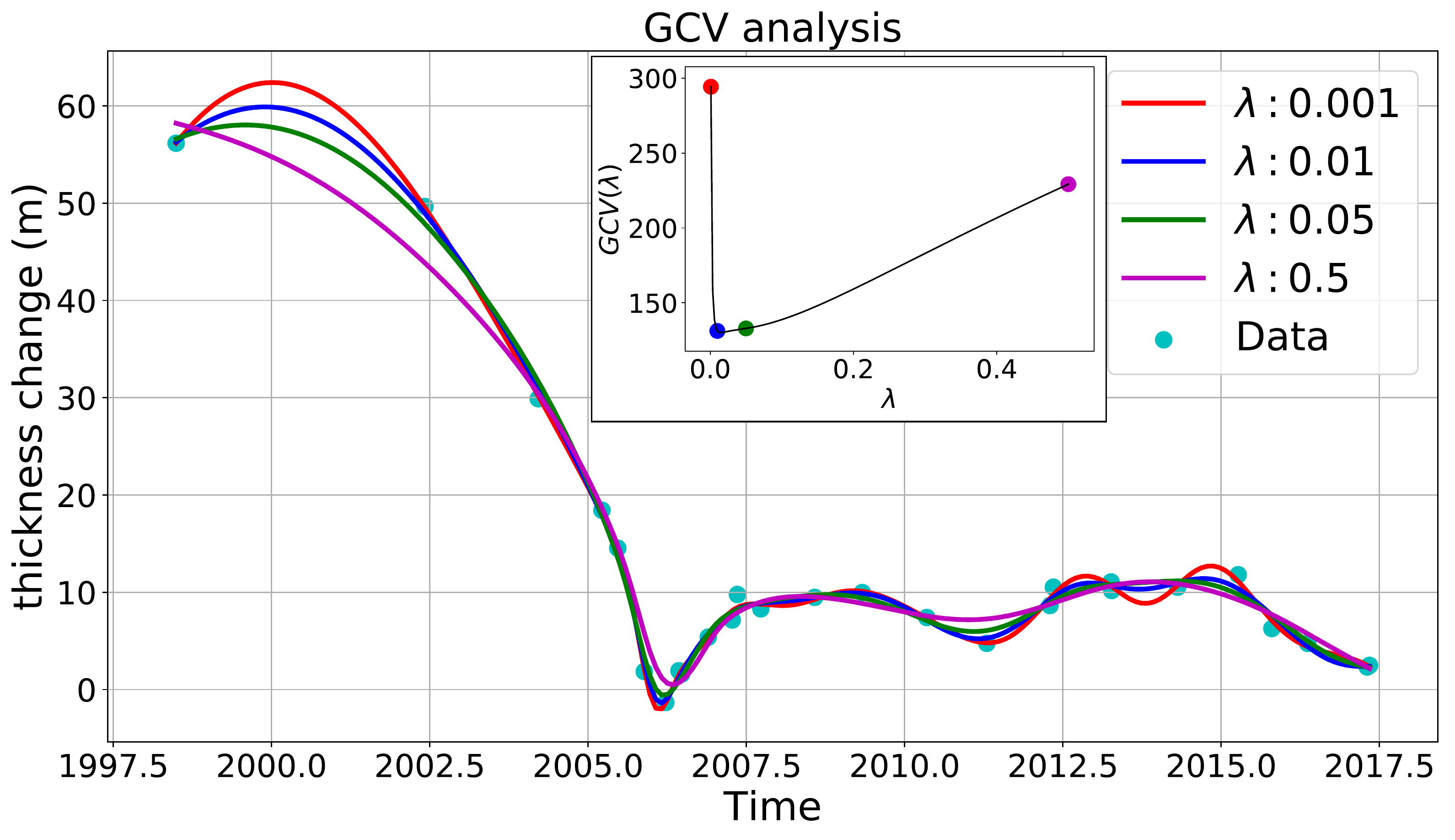}
\caption{ALPS prediction of time series TS:1 with different smoothing parameters ($\lambda$) (other hyperparameters fixed).  The GCV estimates corresponding to $\lambda = 0.5$ (magenta) and $\lambda = 0.001$ (red) are both high, and the corresponding solutions are underfitting and overfitting the data respectively. Models with  $\lambda = 0.01$ and  $\lambda = 0.005$ correspond to low GCV estimate and provides overall good fit. Based on minimum GCV, ALPS selects the solution corresponding to $\lambda = 0.01$.}
\label{gcv}
\end{figure}

\subsection{Estimation through ALPS}\label{fittingpspline}

To obtain prediction from ALPS, several hyperparameters must be defined. These include, 1: Knot distribution,  2: Number of knots, 3: Degree of the basis functions (p), 4: Order of penalty (q) and, 5: Smoothing parameter  $\lambda$ (\ref{lammb}), needed to smooth the fit. In this section we explain the Generalized Cross Validation (GCV) criterion implemented by ALPS that forms the foundation for selecting and determining these hyperparameters.  GCV allows for a tight fit to  the data and usually avoids matrix conditioning issues while training the model, which is important for irregularly sampled observations. Below we provide a basic overview of the GCV statistic and refer to \cite{ruppert2003semiparametric}, \cite{wahba1990spline} and \cite{craven1978smoothing} for more details. GCV statistic quantifies the generalization cost and aims to find a model which strikes the best compromise between model complexity and inference capabilities among the given choices  of models (models here refer to different approximations produced by different combinations of hyperparameters). The GCV statistic is given as

\begin{equation}
GCV(\lambda) = \sum_{i=1}^n {\bigg( \frac{\{(I - H)y\}_i }{1- n^{-1}tr(H)}   \bigg)}^2
\label{GCV_eq}
\end{equation}

with n being the number of data points in the time series and y  the response variable. H is known as the smoother matrix, defined as

\begin{equation}
H = B(B^T B + P)^{-1} B^T
\end{equation}

and $tr(H)$ is the trace operator acting on $H$. B represents the basis function matrix. The model with the least GCV statistic value is regarded as the best model. \cite{ruppert2003semiparametric} provides a detailed background on this. The example  in Fig. \ref{gcv} illustrates the selection of $\lambda$ based on the GCV statistic, showing the minimum GCV is associated with value of $\lambda$ that provides the best generalization (best compromise between overfitting and underfitting). Now we provide the different criteria used by ALPS to select the different  hyperparameters. Towards the end of this section, we bring everything together by explaining the overall ALPS algorithm.

\subsubsection{Knot Distribution}\label{knot}

For defining the distribution of basis functions,  the number and  location of the knots need to be selected. With $m$ sections, the number of knots become $m+1$. Next we infer the distribution of these knots. Following the recommendations of \cite{ruppert2003semiparametric}, we distribute knots based on quantiles of data. The location of the first and the last knot is fixed at the first and last data point, respectively. With $t_j$ as the $j^{th}$ time instance at which an observation is made, the location of the $a^{th}$ knot ($1 \leq a \leq m-1$) is computed as

\begin{equation}\label{quant}
u_a = \bigg( \frac{a}{m} \bigg)^{th} \textrm{ sample quantile of unique }t_j 
\end{equation}
Here `unique $t_j$' gives the discretized set of non-repeating time instances at which observations are available. 

\begin{figure}[h]
\centering
\includegraphics[width=3.5in]{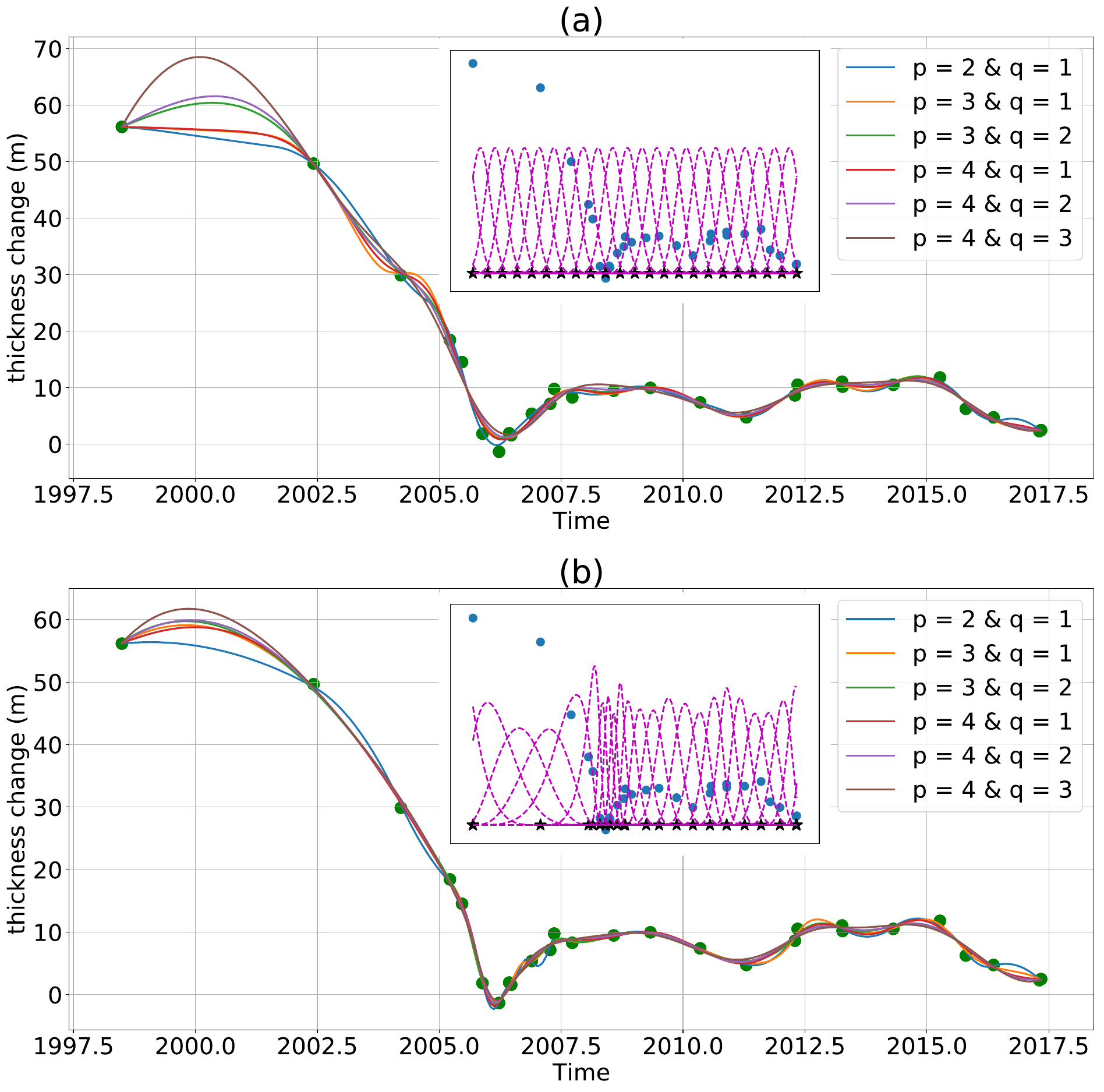}
\caption{(a): Approximations of TS:1 produced by (a) equidistant knots and (b) data-dependent knots; both with optimum number of sections (m) computed through GCV minimization. Insets show  sample distributions of basis functions for p = 4, q = 2. Models are calculated for six different combinations of  basis degree (p) and orders of penalty (q).}
\label{new_3}
\end{figure}

Fig.~\ref{new_3} illustrates the performance of ALPS with equidistant  and data-dependent (quantile-based) knot distributions with different combinations of  degree of basis functions (p) and orders of penalty (q) for modeling TS:1 (time series 1 in Table \ref{timeseries}). The insets show the distribution of the basis functions with uniform knots and data dependent knots in (a) and (b) respectively. The time series data in the insets is the thickness time series, normalized to the magnitude of the basis functions.   Fig. \ref{new_3} is crucial in understanding the sensitivity of the approximation with respect to the distribution of the knots. Firstly, with data-dependent knot distribution, the approximations are  less sensitive to the degree of the basis function and the order of the penalty. This also avoids overfitting as can be seen in Fig.~\ref{new_3}(b) around 2000 and 2004. Secondly,  the rapid reversal of ice thinning to thickening in mid 2005 is captured accurately by the approximation using data-dependent knots, while in the case of equidistant knots the trend reversal is smoothed out.  This is because data-dependent bases have a higher density in regions of dense data (leading to good approximations) and low density in regions of sparse data (avoiding overfitting and promoting generalization).

\subsubsection{Number of Knots}\label{numknots}
We test different number of knots, ranging from 2 to n where n denotes the total number of data points. The  number of knots that produce the smallest GCV statistic is regarded as most suitable. Based on the definition of B-splines  in (\ref{basis}), the number of knots determine the number of basis functions. Hence by choosing the `minimum' number of knots from all the configurations giving minimal GCV cost, we are choosing the optimal model complexity which also produces acceptable approximations and generalizations.

\subsubsection{Degree of basis functions (p) and order of penalty (q)}\label{pq}
\begin{figure}[h]
\centering
\includegraphics[width=3.5in]{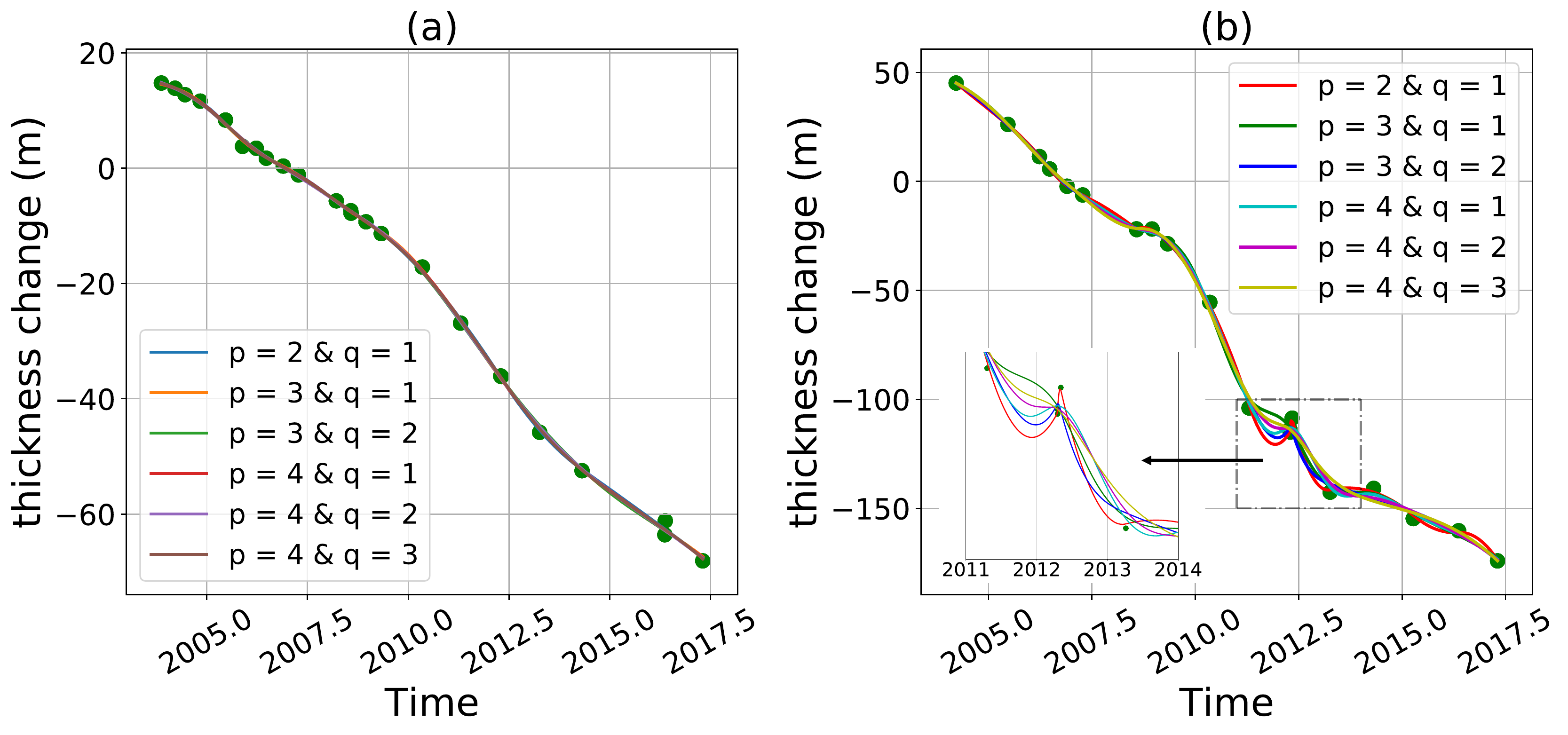}
\caption{Sensitivity of P-spline approximations  to the selection of the degree of basis functions  (p) and the order of penalty (q). Ice thickness change time series on Midgard Glacier, (a) upstream, on higher elevation (TS:2), and (b) near the terminus (TS:3). See Fig.~\ref{Locmap} and Table~\ref{timeseries} for more information on TS:2 and TS:3. Inset in (b) highlights the sensitivity of the approximation to the different selections of p and q in 2011-2014.}
\label{new_4}
\end{figure}

As illustrated in Fig.~\ref{4}(a), the number of knots across which an individual basis function influences the predicted curve is determined by the degree (p) of B-spline basis function ($B_i(t;p)$ is non-zero for $t \in [u_i,u_{i+p+1})$).  Hence for time series smoothing, we consider the order  of the penalty (q) to be less than p. In this study, we only considered $p\leq4$ to avoid loosing the benefits of a local modeling strategy,  as higher degree B-spline basis functions have broader support. Therefore, the only available selections for p  are $p = 4, 3, 2$, which, combined with the possible selections of q ($q < p$), result in the six cases shown in Figs.~\ref{new_3} and \ref{new_4}.

We illustrate the selection of the parameters p and q using examples from the  Helheim (TS:1, Fig.~\ref{new_3}) and Midgard glaciers (TS:2-3, Fig. \ref{new_4}). These neighboring glaciers, both terminating in the Sermilik Fjord in east Greenland (Fig.~\ref{Locmap}), have been   thinning rapidly during since the early 2000s. However,  unlike the rapid variations of thinning and thickening on the Helheim Glacier (e.g., Fig.~\ref{new_3}), Midgard Glacier exhibited a monotonous, gradual thinning (Fig.~\ref{new_4}).   
 For gradual changes and well distribued samples, the choice of p and q is usually not important (e.g., Fig~\ref{new_4}(a)). However, in case of uneven sampling, different p, q selections can provide different approximations and  result in under- or overfitting, such as the case in the data gap around 2000 in TS:1 (Fig.~\ref{new_3}(a)) and during rapid changes in 2011-2014 in TS:3 (Fig.~\ref{new_4}(b)). Based on the  overall good performance, we  selected  p = 4 and q = 2 or 3 for  modeling of ice sheet time series data in this study.

\subsubsection{Confidence bounds}\label{confidence}
 We use t-Confidence Intervals (CI) because they are  suitable for  small data sets and closely mimic the normal distribution bounds  as the size of the data set increases. We estimate the variance of the error term by using the unbiased estimator of $\sigma_{\epsilon}^2$, given in \cite{ruppert2003semiparametric} as
 
\begin{equation}\label{fitvar}
\hat{\sigma_{\epsilon}}^2 = \frac{{||y - B \hat{\Theta}||}^2}{df_{res}}
\end{equation}

with the degree of freedom of residual ($df_{res}$) for the non-parametric case (\cite{ruppert2003semiparametric}) given as

\begin{equation}\label{dfres}
df_{res} = n - 2 tr(H_{\hat{\lambda}}) + tr(H_{\hat{\lambda}} H_{\hat{\lambda}}^T)
\end{equation}

where $H_{\hat{\lambda}}$ is the smoother matrix at $\hat{\lambda}$. Following the recommendations of \cite{ruppert2003semiparametric}, the standard deviation of the error term can be estimated as follows

\begin{equation}
\widehat{st.dev}(\hat{f}(t) - f(t)) = \hat{\sigma_{\epsilon}} \sqrt{ B_t (B^TB + \hat{P})^{-1}B_t^T}
\label{gcvbound}
\end{equation}

where  $B_t$ represents the basis functions at the prediction time instances and B represents the basis function values at the observed time instances. These CIs are the same as the ones proposed by \cite{wahba1983bayesian} and \cite{nychka1988bayesian} from a Bayesian perspective. Finally, the $100(1-\alpha)$ \% CIs for $f(t)$ is

\begin{equation}
\hat{f}(t) \pm t\bigg(1 - \frac{\alpha}{2}; df_{res} \bigg) \widehat{st.dev}(\hat{f}(t) - f(t))
\label{confbounds}
\end{equation}

\subsubsection{Derivative Estimation}
\label{derivative}

The estimation of rates of change of $dh$ can be done following  \cite{ruppert2003semiparametric}. Hence for models in (\ref{modelgcv}) the corresponding derivatives of the mean function are

\begin{equation}
\hat{f}^{'} (t) = B_t^{'} \hat{\Theta}
\end{equation}

For getting the derivatives of the basis functions $B_t$ ($B_t^{'}$) we refer \cite{piegl2012nurbs} and obtain the formula for the 1st order derivative of the $i^{th}$ B spline basis of degree p

\begin{equation}
B_{i}^{'}(t;p) = \frac{p}{t_{i+p} - t_i} B_{i}(t;p-1) - \frac{p}{t_{i+p+1} - t_{i+1}}B_{i+1}(t;p-1)
\label{derv_bases}
\end{equation}

Then following the recommendation of \cite{ruppert2003semiparametric}, we use the parameters estimated by GCV and get the first derivative of the fitted curve.

\IncMargin{1em}
\begin{algorithm}
\SetKwData{Left}{left}\SetKwData{This}{this}\SetKwData{Up}{up}
\SetKwFunction{Union}{Union}\SetKwFunction{FindCompress}{FindCompress}
\SetKwInOut{Input}{Input}\SetKwInOut{Output}{Output}
\Input{Time Series data: $D = [T,y]\in \R^{n \times 2}$\\
Degree and Penalty: $p,q$\\
Prediction time instances: $T_t$\\
Confidence level: $\alpha$
}
\Output{Prediction with CI: $\hat{f}(t)$, $std_1$ \\
$1^{st}$ derivative with CI: $\hat{f}^{'}(t)$, $std_2$
\emph{\noindent\rule{8.1cm}{0.4pt}}
}
\emph{Initialize: $m =1, TH=0$}\\
\While{$m < n$}{
Number of basis functions: $c \leftarrow m+p$\\
$U \leftarrow Knot\_vector(D,p,m)$\\
$b \leftarrow Basis\_functions(m,p,U,D)$\\
$[\lambda, cost] \leftarrow min\_GCV(D,b,q,c)$\\
\If{$cost < TH$ or $m == 1$}{
$TH \leftarrow cost$\\
$[\hat{\lambda}, \hat{m},\hat{U}, B] \leftarrow [\lambda, m,U, b]$\\
}
$m \leftarrow m+1$
}
$\hat{c} \leftarrow \hat{m} + p$\\
$\hat{P} \leftarrow Penalty(q,\hat{c},\hat{\lambda})$\\
$\hat{\Theta} \leftarrow ({B}^T{B} + \hat{P})^{-1}{B}^Ty$\\
$B_t \leftarrow Basis\_functions(\hat{m},p,\hat{U},T_t)$\\
$B^{'}_t \leftarrow Basis\_functions\_derv(\hat{m},p,\hat{U},T_t)$\\
$[\hat{f}(t),std_1] \leftarrow pred(D,{B},B_t,\hat{\Theta},\hat{P},\alpha)$\\
$[\hat{f}^{'}(t),std_2] \leftarrow pred\_derv(D,{B},B_t,B_t^{'},\hat{\Theta},\hat{P}, \alpha)$\\
return $[\hat{f}(t),std_1,\hat{f}^{'}(t),std_2]$
\caption{{ALPS}}\label{Algo1}
\end{algorithm}\DecMargin{1em}

In order to obtain the CIs for the derivatives, we again follow the approach from \cite{ruppert2003semiparametric} and get the following bounds

\begin{equation}
    \widehat{st.dev}(\hat{f}^{'}(t) - f^{'}(t)) = \hat{\sigma_{\epsilon}} \sqrt{ B_t^{'} (B^TB + \hat{P})^{-1}{B_t^{'}}^T}
    \label{gcvbound2}
\end{equation}

We have again used $t(1- \frac{\alpha}{2}, df_{res})$ scaling of the appropriate standard deviation to get $100(1-\alpha)\%$ confidence bounds. A detailed application of derivative computation and `knowledge inference' has been included in Section \ref{helheim}.

\subsection{ALPS algorithm}\label{implementation}
The steps involved in the ALPS algorithm are shown in Algorithm \ref{Algo1}. The algorithm takes data set D, containing $n$ samples of \{time, height\} pairs, as  input. The default values of hyperparameters p and q are set to  p = 4 and q = 2. However,   other degree and penalty order values can also be selected.  $T_t$ are the time instances at which we want to make predictions and determine the confidence intervals. $\hat{f}(t)$ is the mean prediction with the corresponding confidence interval $std_1$. Similarly $\hat{f}^{'}(t)$ and $std_2$ are the estimates of the first derivative of the prediction.

\begin{figure}[h]
\centering
\includegraphics[width=3.5in]{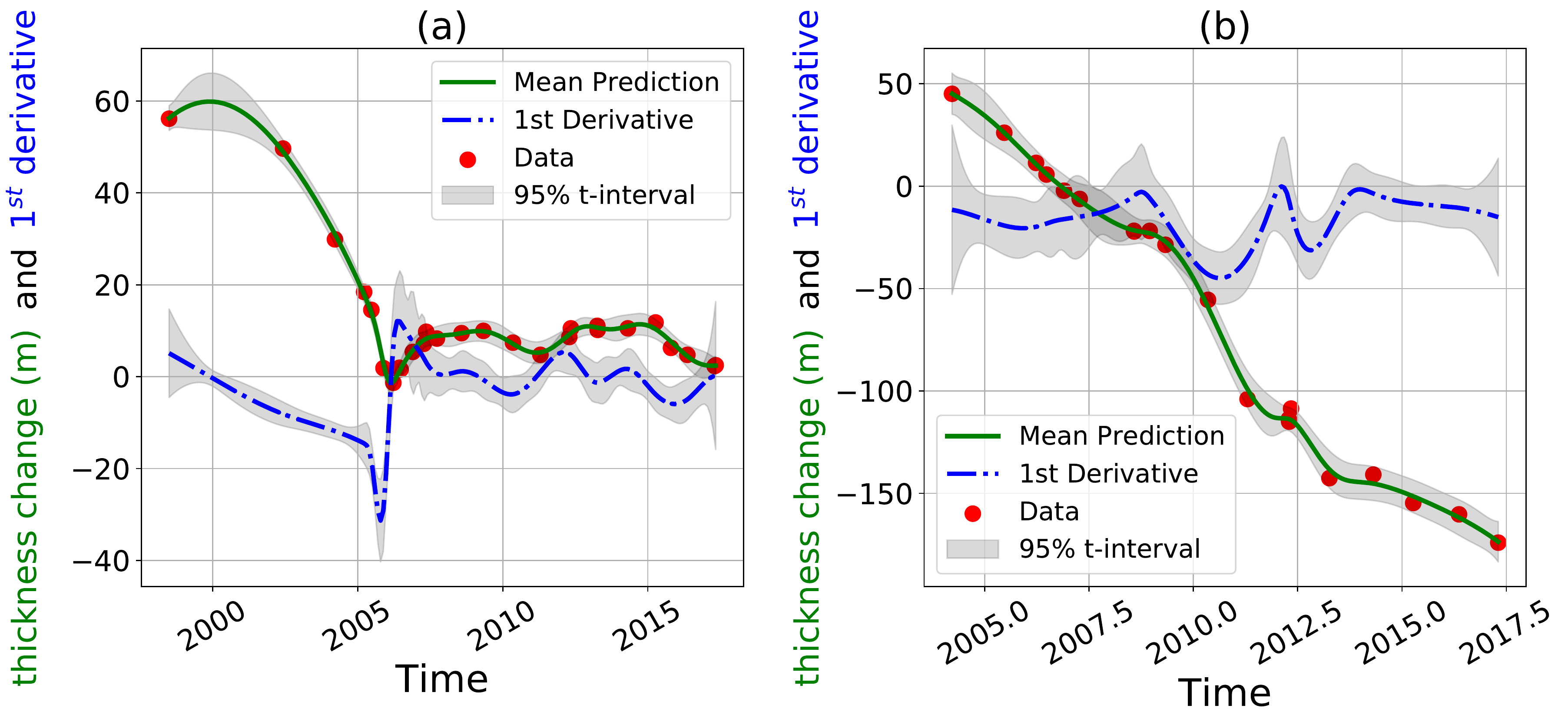}
\caption{ALPS output (mean prediction and $1^{st}$ derivative) for TS:1 (a) and TS:3 (b) calculated using p=4 and q=2. Grey bands show 95\% t-CI.}
\label{new_5}
\end{figure}

Representing the proposed total number of sections on the predicted curve by m, we start with m = 1 (one knot each at the beginning and end with no intermediate knots) and  compute the knot vector U (as described in section \ref{knot}), then with computed bases b as per (\ref{basis}) (here we are using a temporary notation b for basis functions and the basis functions with optimal number of knots will be represented as B), we compute the value of $\lambda$ that minimizes the GCV metric (\ref{GCV_eq}). In line 6 of Algorithm \ref{Algo1}, $\lambda$ represents this optimal value with cost representing the corresponding minimized $GCV$ cost. If this cost is less than the cost of any previous number of sections ($m$) considered, then we save the details for the current $m$. Then using the details related to optimal m ($\hat{m}$), we compute $\Theta$ as in (\ref{theta}). Then, considering the time instances $T_t$, at which predictions are required, we compute the bases $B_t$ (again using (\ref{basis})) and derivative of the bases $B_t^{'}$ (using (\ref{derv_bases})). Then predictions $\hat{f}(t)$ and $\hat{f}^{'}(t)$ are computed as $B_t \hat{\Theta}$ and $B_t^{'} \hat{\Theta}$ respectively. The corresponding confidence intervals ($std_1$ and $std_2$) are computed using eq. (\ref{gcvbound}) and eq. (\ref{gcvbound2}) respectively. These computations have been represented by subroutines $pred$ and $pred\_derv$ in Algorithm \ref{Algo1}. Fig.~\ref{new_5} demonstrates, ALPS  provides good approximations   even for difficult time series, for example, with  abrupt changes (TS:1, Fig.~\ref{new_5}(a)) or  non-uniform sampling (TS:3, Fig.~\ref{new_5}(b)).

\subsection{Outlier Detection}\label{out}
Besides modeling the time series, we can also use ALPS for detecting outliers. The algorithm is based on a two-level thresholding strategy with user-chosen thresholds. A data point is identified as an outlier when it is outside the 99\% t-confidence scaled by $threshold_1$. After the removal of the outliers, the model is again fitted following the same strategy with $threshold_2$ scaling of the standard error bound. The two-level strategy is effective because of the sensitivity of the least-squares cost functions  to outliers. From our experience, a value of 3 for $threshold_1$  and 1.2 for $threshold_2$ works well for the datasets under consideration.

\begin{figure}
\centering
\includegraphics[width=3.5in]{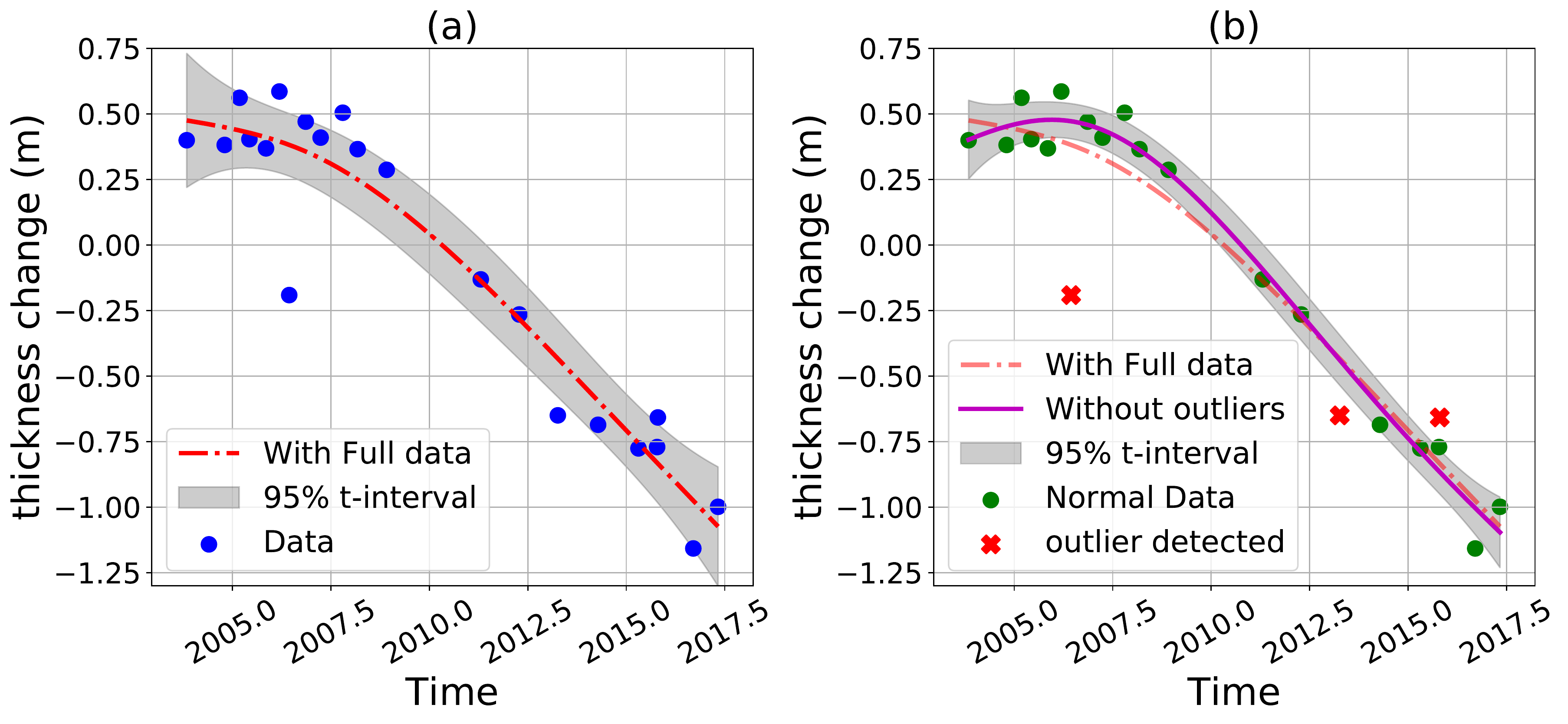}
\caption{Outlier detection and removal by thresholding using a 99\% t-CI derived by the ALPS (TS:4). (a)  ALPS prediction by using all data points; (b)  approximation after removing three  outliers identified by the two-level approach discussed in Section \ref{out}.}
\label{new_6}
\end{figure}

Fig.~\ref{new_6} illustrates the performance of the outlier detection algorithm on a time series (TS:4), which depicts slow dynamic thinning in a region south of the Jakobshavn Glacier. The model in Fig.~\ref{new_6}(a) is based on all the data points. However,  applying the two-level outlier detection, one data point is detected as an outlier at the first level and two others  at the second level (shown with `x' markers (Fig.~\ref{new_6}(b)). Note the improvement in the quality of fit and narrower confidence bounds when the outliers are removed  (Fig.~\ref{new_6}(b)).

\subsection{Comparison with other methods}\label{comp_method}

\begin{figure}[h]
\centering
\includegraphics[width=3.3in]{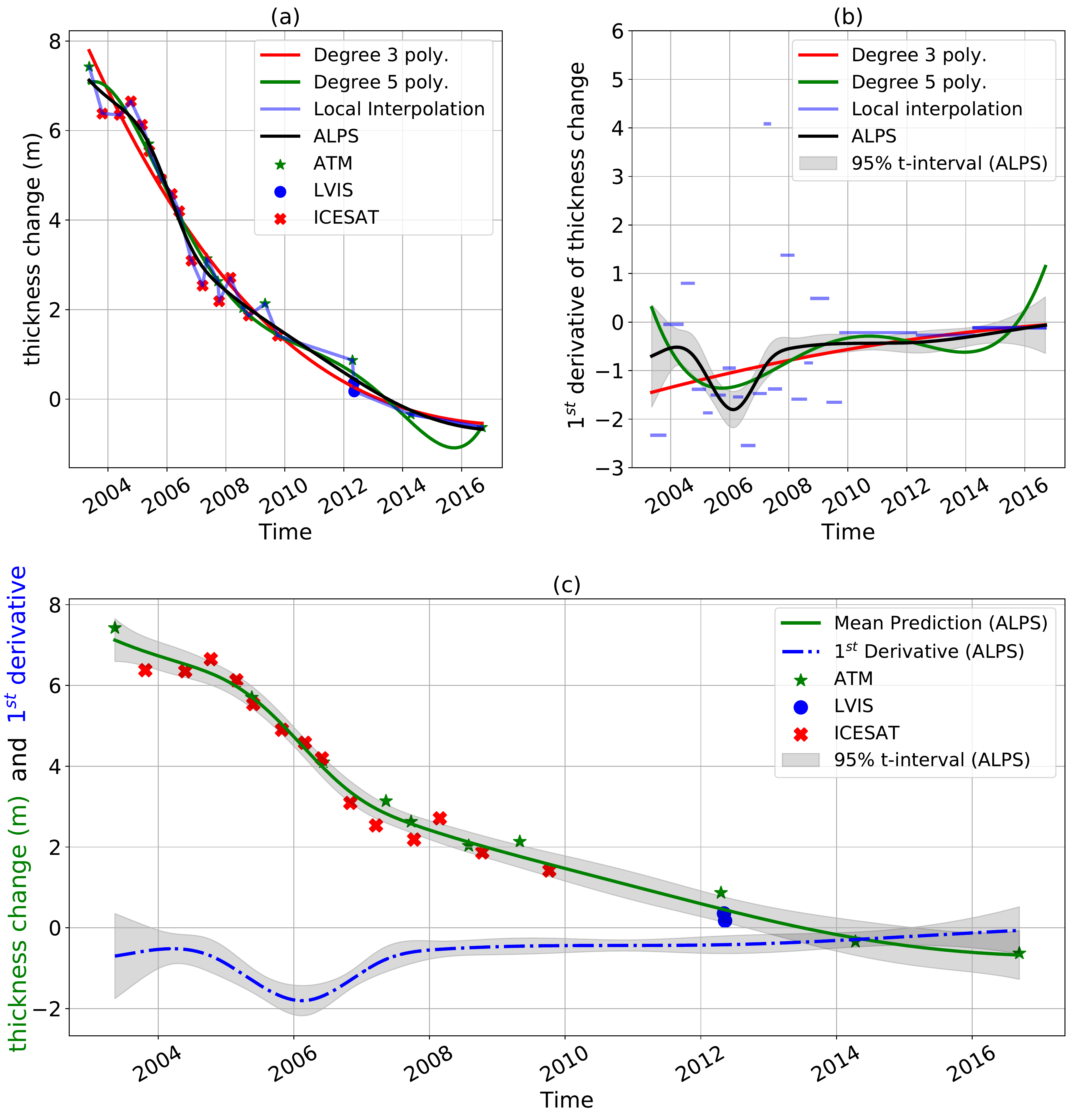}
\caption{Comparison of ALPS (on TS:0) with local interpolation and global polynomial models of degree 3 and 5. (a) shows the comparison of predictions, (b) shows the corresponding first derivatives and (c) individually shows the performance of ALPS on TS:0 (with confidence intervals for mean prediction and 1$^{st}$ derivative).}
\label{comparison2}
\end{figure}

In this section, we compare the predictions made by ALPS with a few other models studied in Fig. \ref{comparison1}. The analysis is shown in Fig. \ref{comparison2}. For global approaches we are now considering polynomials of degree 3 (found to work relatively well on the considered time series in Fig. \ref{comparison1}) and degree 5 (as we wanted to compare ALPS with a higher degree polynomial). Regarding local approaches, we have chosen to compare to local linear interpolation. Owing to the unavailability of predictions in most of the time windows, local approximation with half annual resolution is not considered here. Starting with the results shown in Fig. \ref{comparison2}(a), we compare mean prediction produced by ALPS with the results from other considered models. Considering performance with respect to global approaches, ALPS performs better than degree 3 polynomials, which is clearly oversmoothing the prediction from 2004 to 2006. Also from 2007 to 2009, the predictions from degree 3 polynomial are showing a biased behavior. Besides tackling these issues of oversmoothing and biased fitting, ALPS is also able to avoid any spurious fluctuations, typical of a model with high degrees of freedom. Such oscillatory behavior is demonstrated by degree 5 polynomial from 2014 to 2016 in Fig. \ref{comparison2}(a). Regarding local models, the superiority of ALPS over local interpolation is evident from the first derivatives shown in \ref{comparison2}(b). The noisy nature of the derivative (for local interpolation) with sharp and big jumps in \ref{comparison2}(b) clearly shows that the model is fitting to the noise in data as processes in nature are usually not that erratic. Similar to the mean prediction case, even the derivatives (of polynomial models) in (b) show that ALPS avoid rapid spurious fluctuations (like degree 5 polynomial) while learning the structure in data without oversmoothing (like degree 3 polynomial). For showing the performance of ALPS on TS:0 more clearly, we have dedicated panel (c) in Fig. \ref{comparison2}. In this panel, besides showing the predictions, we also show the confidence intervals.

\begin{figure}
\centering
\includegraphics[width=3.3in]{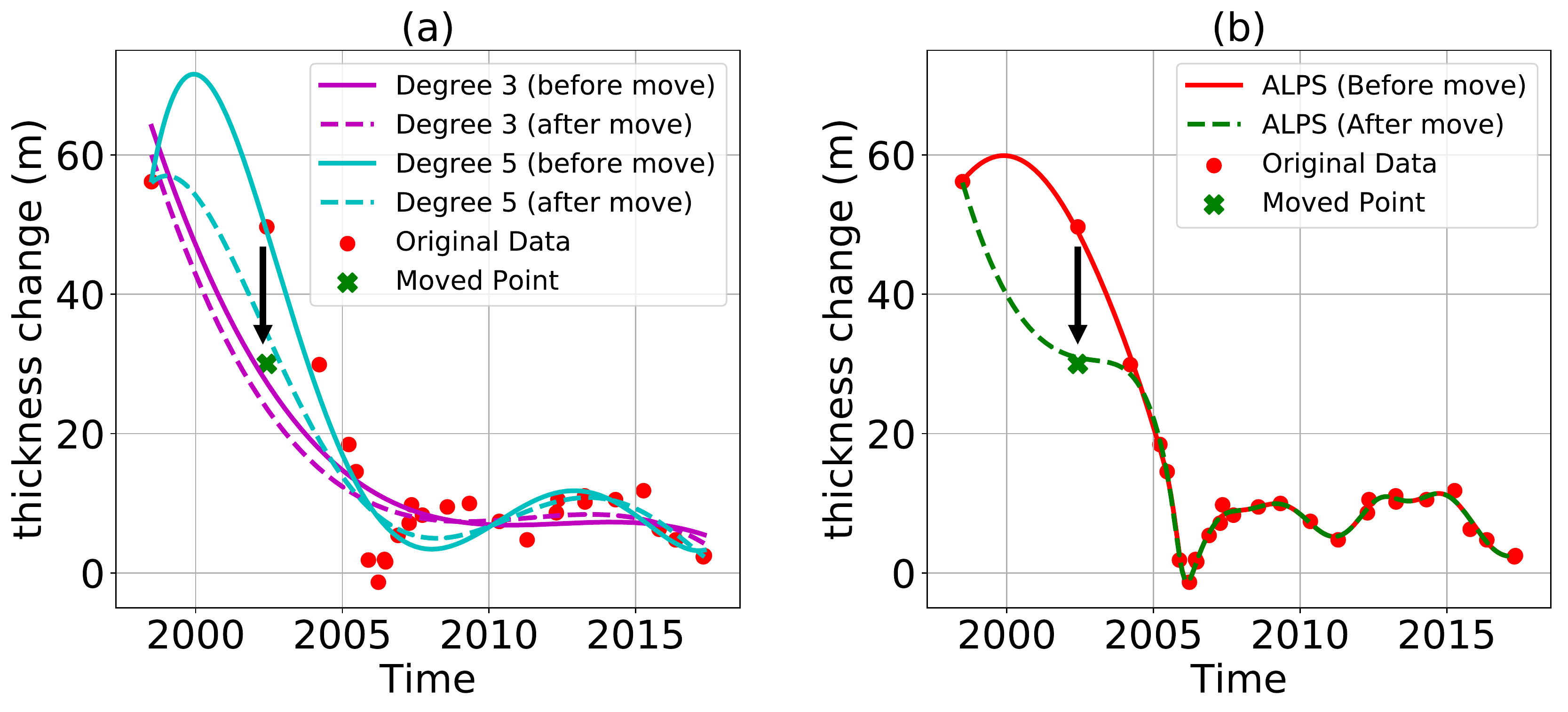}
\caption{The impact of a local perturbation (caused by moving a single data point) on predictions with (a) degree 3 and degree 5 polynomial; (b) ALPS on TS:1. The localized sensitivity of the ALPS based approximation in (b) shows its robustness to local noise.}
\label{new_2}
\end{figure}

Moving forward, we analyze the sensitivity of ALPS and polynomial models to changes in data (specifically one data point for our case). Fig.~\ref{new_2} illustrates this result. Due to global basis functions polynomial approximations exhibit a global sensitivity to local perturbations, i.e., even one erroneous observation (due to  a measurement or processing error) could  result in significant prediction errors over the whole time span (Fig.~\ref{new_2}(a)). The ALPS prediction, however, is only affected locally by a single outlier data point, as  Fig.~\ref{new_2}(b) demonstrates. This analysis further supports out claim that ALPS provides most generalizable (avoides overfitting and underfitting) predictions with high robustness and being only locally sensitive to outliers.

\section{Applications}\label{applications}


\subsection{Combining laser altimetry measurements 
with a Firn Densification Model (FDM) to derive high resolution ice surface elevation change time series}\label{firn}


The elevation of the ice sheet surface  continuously changes due to physical processes acting on the surface (e.g., melt, accumulation, sublimation), within the firn column (e.g., firn compaction, ice lense formation), at the base (e.g., basal melt) and  to the  vertical deformation of the underlying earth crust \cite{Cuffey:2010}.   

To facilitate the computation of a high-temporal resolution surface elevation change time series,  we partition the ice sheet surface elevation change ($h$)  into a rapidly changing component caused by surface processes ($h_{s}$) and a slowly changing component ($h_{dibc}$) that includes changes due to ice dynamics (d), internal (i) and basal (b) processes, vertical crustal deformation (c)    \cite{csatho:2014jja}: 
  
\begin{equation}
   h(t)= h_s(t)+h_{dibc}(t)
   \label{partitionh}
\end{equation}

 The sparse and uneven measurements of ice sheet surface elevations prevents the derivation of a high-temporal resolution record directly from the observations.  However, surface elevation change time series caused by surface processes ($h_{s}$) are   available with dense temporal sampling (5-10 days) from Firn Densification Models (FDM) forced by the outputs of a Regional Climate Models (RCMs) or Earth System Models (ESMs) (e.g., \cite{Lenaerts:2019if}). Note that following the terminology of recent publications (e.g., \cite{KuipersMunneke:2015fh, Smith:2020en},  we refer to the total vertical velocity of ice sheet surface due to firn and SMB processes as FDM  ($h_{s}$).

Using (\ref{partitionh}), $h_{dibc}$  can be calculated as the difference between  surface elevations (modeled by ALPS from altimetry observations, $h$) and  surface process related changes (modeled by RCMs or EMSs, $h_{s}$).  Since $h_{dibc}$ represents the low-frequency component of the ice sheet elevation change, it can be modeled even from a sparse data set. Our desired final result, ($h(t)$), a high-temporal resolution  surface elevation time series can be obtained by combining  the high resolution  $h_{s}$ and the modeled  $h_{dibc}$.   

\begin{figure}[h]
\centering
\includegraphics[width=3.3in]{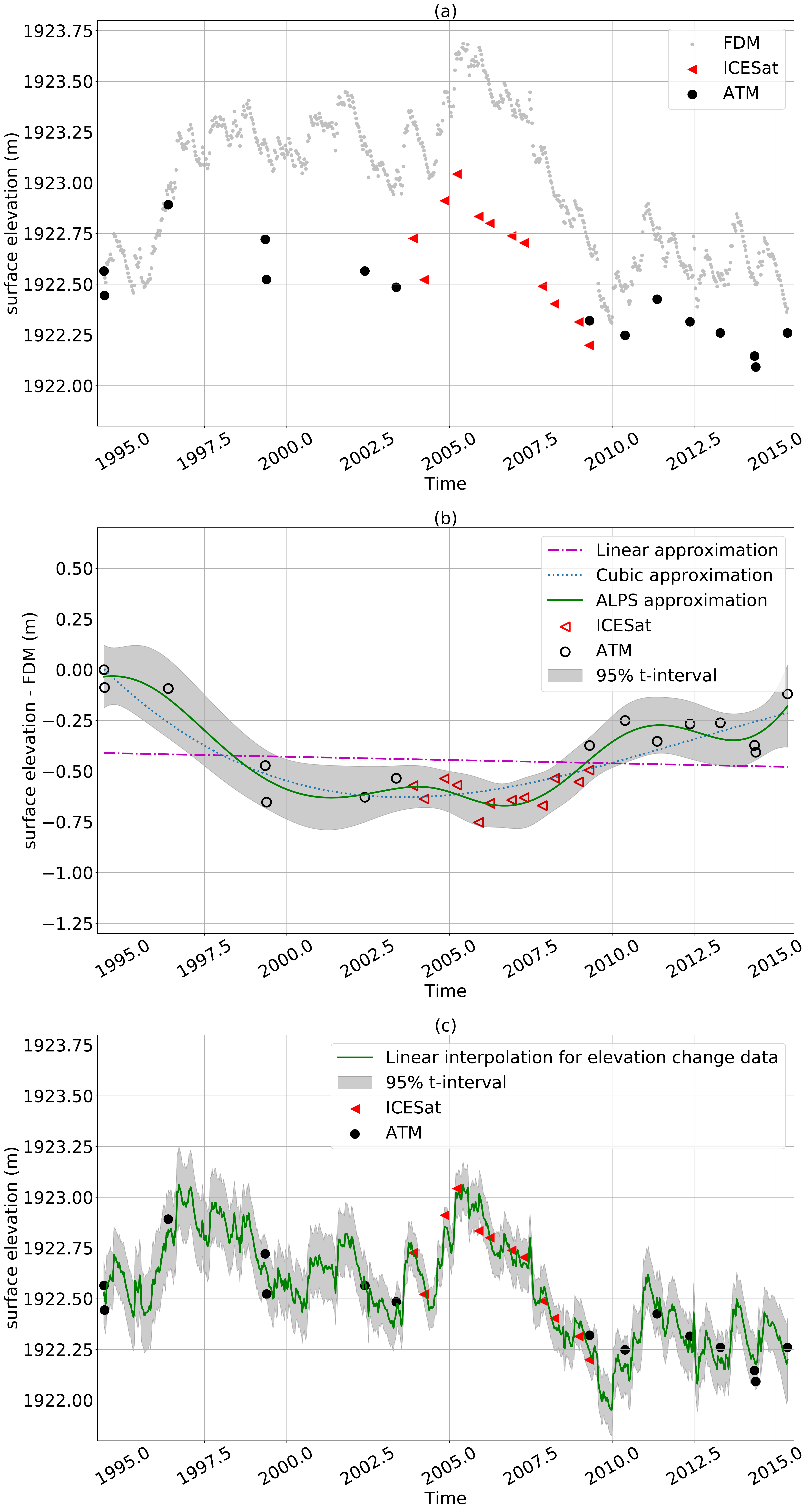}
\caption{Calculation of a high temporal resolution ice sheet surface elevation  time series, TS:5 in NW Greenland, Site 9 in \cite{KuipersMunneke:2015fh}. (a) Ice sheet elevations reconstructed by SERAC (solid traingles:  ICESat; solid circles: ATM) and FDM from RACMO2.3p2 (smaller solid circles). (b) Difference between ice  sheet elevation and FDM (empty circles and triangle markers) with linear, cubic polynomial and ALPS modeling. Shaded band shows 95 $\%$ CI for ALPS. (c) High temporal resolution ice surface elevation  time series (solid line with  95 $\%$ CI), computed as the sum of the FDM model and the ALPS approximation of the difference component.}
\label{app1}
\end{figure}

Fig.~\ref{app1} illustrates the calculation of high temporal resolution ice sheet elevation time series  from laser altimetry measurements with an example from the interior region of the ice sheet in NW Greenland (TS:5 , Site 9 in \cite{KuipersMunneke:2015fh}). 
 Here $h(t)$ is reconstructed by SERAC from NASA's ATM and ICESat laser altimetry measurements collected between 1993 and 2017 (larger solid circle and triangle markers in Fig.~\ref{app1}(a)). $h_s(t)$ is  a 10-day resolution  FDM record spanning 1960-2016, which was simulated using the IMAU-FDM firn model  forced by the output of the RACMO2.3p2 regional climate model  \cite{Ligtenberg:2018}. It is shifted to the first altimetry measurement (Fig.~\ref{app1}(a), smaller solid circles).

The difference between the measured ice sheet elevation change ($h(t)$) and the modeled surface processes driven elevation change  ($h_{s}(t)$) is shown by hollow circle and triangle markers in Fig.~\ref{app1}(b).  We applied three different models (linear, cubic and ALPS) to model this data set.  The P-spline based model of ALPS provided a more accurate fit than the polynomial models over the entire domain. Therefore, to produce an interpolated high temporal resolution ice sheet elevation time series (Fig.~\ref{app1}(c)), we combine the ALPS approximation with the FDM model at 10-day resolution.  The ALPS-based approximation, which uses GCV model fitting, was computed with a B-spline bases of degree (p) 4 and penalty (q) 2.  Note that the error of  $h_{dibc}$ reflects the modeling error only, i.e., we assume that the systematic errors of $h$ and $h_{s}$ are negligible and all errors are non-correlated. This assumption might result an underestimation of the error in  $h_{dibc}$, but has no impact on the error of the ice surface elevation.

\subsection{Investigating outlet glaciers changes, a case study of  Helheim Glacier} \label{helheim}

While the proposed penalized spline based approach was originally developed to approximate discrete time series of ice sheet elevation and thickness changes, it is also suitable for modelling other time series, such as point-based reconstruction of ice velocity changes (e.g., \cite{Rosenau:2015bg}) or area-based reconstruction of calving-front locations (e.g., \cite{Schild:2013ie}). Here we demonstrate the efficiency of ALPS to generate ice sheet change records from a variety of observations  for investigating the  dynamics of Helheim Glacier in East Greenland.

 After several decades of equilibrium when ice sheet gains roughly equaled losses, the Greenland Ice Sheet has started to loose mass with increasing rates in the late 1990s  (\cite{theIMBIEteam:2019, Mouginot:coa}). About 48$\%$ of this total loss in 1992-2018 is attributed to increasing discharge from outlet glaciers. While outlet glacier acceleration, thinning, and retreat has been widespread, the response of individual glaciers to climate forcing varies both in magnitude and timing \cite{csatho:2014jja}. This indicates that the effect of climate forcing  is modulated by local conditions, such as bed geometry and ocean conditions. With the advent of remote sensing methods, detailed, accurate  records of outlet glacier changes (i.e., dynamic thickness, velocity, and terminus changes) are becoming available  to investigate the timing and propagation of these changes (e.g., \cite{csatho:2014jja,Rosenau:2015bg, Joughin:2019}). However, the fusion of multisensor data remains a challenge. 

We apply   ALPS to investigate the behavior of Helheim Glacier during the accelerated mass loss of the Greeland Ice Sheet. Helheim Glacier is a fast flowing  tidewater glacier that drains $4\%$ of the Greenland Ice Sheet (e.g., \cite{Nick:2014jp}). From 2000 to 2005 Helheim Glacier experienced rapid dynamic mass loss and  thinning due to flow acceleration  (\cite{Bevan:2012hv, Khan:2014db}). The spatiotemporal pattern of the changes initially was consistent with a decrease of resistive forces at the glacier terminus (due to increased calving or reduced sea ice cover, for example) causing  thinning and acceleration  propagating upglacier \cite{Nick:2014jp}. While on most glaciers (e.g., Jakobshavn Isbr{\ae}) thinning and acceleration continue for several years and gradually decrease as the glacier reaches a new equilibrium (e.g., \cite{csatho:2014jja}),  on Helheim Glacier, the initial thinning and accelaration rapidly diminished by 2005. The following re-stabilization period  was associated with short-term periods (1-2 yr) of speedup-slowdown and retreat-advance (\cite{Bevan:2015du, Kehrl:2016}). 

\begin{figure}[h]
\centering
\includegraphics[width=3.3in]{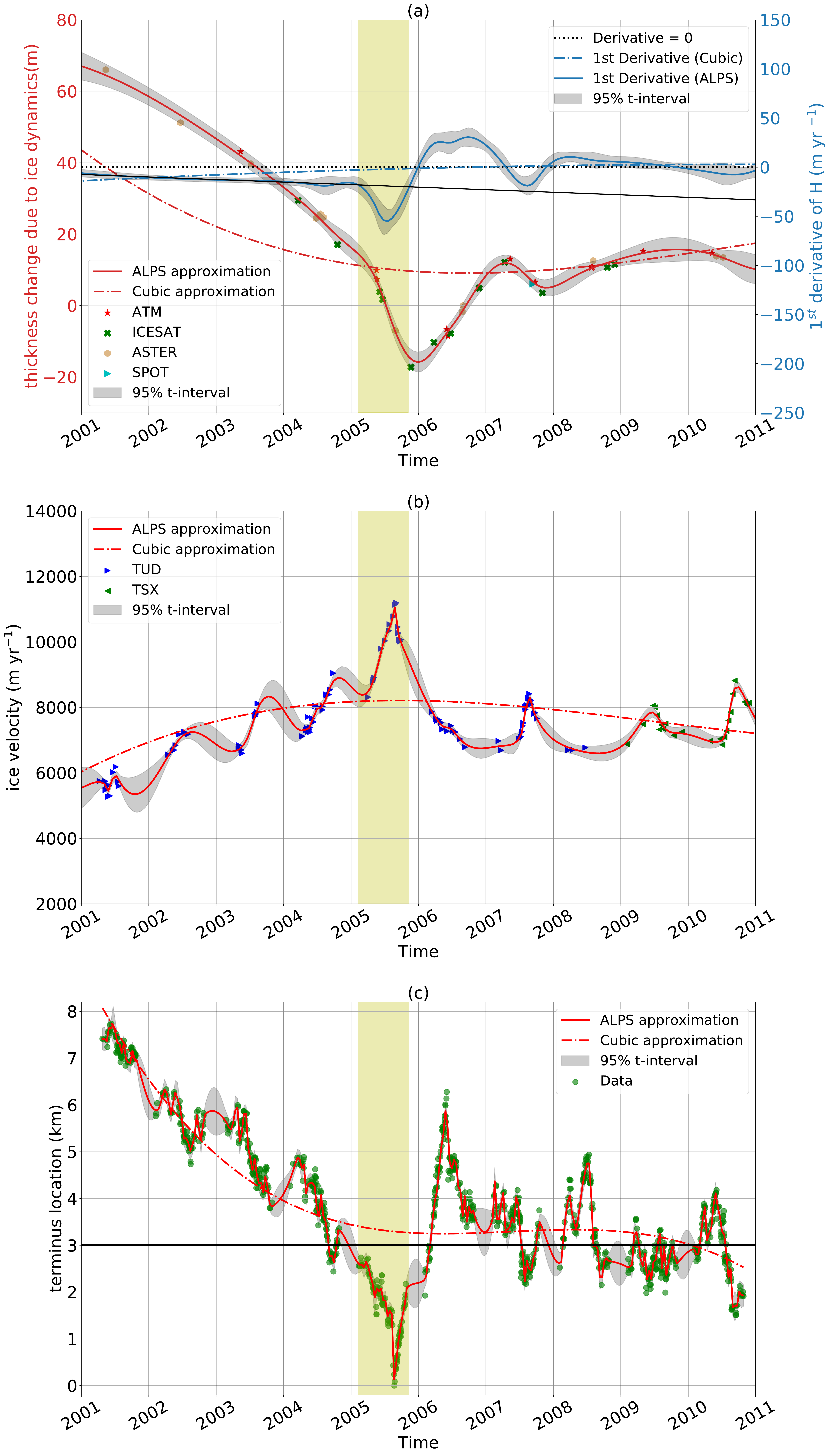}
\caption{Time series of Helheim glacier changes in 2001-2010. (a) Ice thickness change due to ice dynamics referenced to August 31, 2006 from laser altimetry (ATM, ICESat) and stereo DEMs (ASTER, SPOT), with ALPS approximation (red line) and its derivative (blue line) and polynomial approximation with its derivative (dotted lines), (TS:6a). (b) Ice velocity observations from repeat Landsat imagery (TUD) and InSAR measurements (TSX) with ALPS approximation  (solid red line) and polynomial approximation (dotted line), (TS:6b).  (c) Mean terminus distance  normalized to farthest upstream retreat in August 2005 from MODIS imagery with ALPS approximation (red line), and polynomial approximation (dotted line) (TS:6c). See Fig.~\ref{Locmap} for time series locations, Table~\ref{timeseries} and Appendix C for data set references and  details. Grey bands mark 95 $\%$ CIs of ALPS approximations. Yellow band highlights February-November 2005, when  thinning rates exceeded initial 2001-2004 linear trend (black line).}
\label{app21}
\end{figure}

We combined remotely sensed data from several sensors  to generate a detailed record of the glacier for the 2001-2010 period (Fig.~\ref{app21}, \cite{Roberts:2019}). The data sources are listed in Table I and details are provided in Section~\ref{Data}.  

The best altimetry coverage on the main trunk of the  glacier is on the medial moraine at the confluence of its main branches (Fig.~\ref{Locmap}, TS:6a), where two ICESat ground tracks, as well as several ATM swath intersect each other. Because of the poor quality of velocity data sets at this location, we extracted the velocity time series at TS:6b, near 2018 terminus location (Fig.~\ref{Locmap}). As shown by  \cite{Roberts:2019}, 
elevation changes are very similar over the entire main trunk and therefore the two time series could be jointly interpreted. Fig.~\ref{app21}(a) shows the time series of dynamic thickness change  variation at TS:6a. The laser altimetry time series was supplemented by  DEMs derived  ASTER and SPOT-5 and corrected by SERAC  (\cite{Roberts:2019}, Section~\ref{Data}). The velocity change time series  in Fig.~\ref{app21}(b) combines data from two freely available archives, and includes velocities derived from repeat Landsat imagery in 2001-2008 \cite{Rosenau:2015bg} and TerraSAR-X/TanDEM-X  (TSX) SAR imagery in 2009-11 \cite{Joughin:2019}. Finally, terminus positions (Fig.~\ref{app21}(c)) were derived from Moderate Resolution Imaging Spectrometer (MODIS) satellite imagery by Schild and Hamilton \cite{Schild:2013ie}.

The ALPS approximations  of the time series   reveal three major  stages of dynamic changes with distinctly different glacier behavior. We established the timing of the three stages from the dynamic thinning  and its derivative, and consulted the velocity and calving front changes for   interpreting the glacier behavior during the different stages. 

\subsubsection{2001-2004: retreat, speed-up and accelerating dynamic thinning:} During this period all three parameters (thinning, speed-up and retreat) exhibited a linear trend.  Superimposed on the trend, there was a  seasonal velocity signal of mid-summer speed-up correlating with a  terminus retreat  (Fig.~\ref{app21}(b-c),  \cite{Bevan:2015du}). According to \cite{Moon:2014jr}, this pattern, combined with ice velocities that remain relatively high until late winter or early spring indicates that the glacier's dynamic behavior is controlled by melt water availability and terminus retreat.

\subsubsection{2005: rapid dynamic thinning, continuing speed-up and rapid retreat to most inland position, followed by re-advance and  slow-down}
A rapid dynamic event started in early 2005 when dynamic thinning rates became larger than the linear trend of 2001-2004   (yellow band, Fig.~\ref{app21}(a)).
By late August, Helheim Glacier   reached its peak speed of 11000~m~yr$^{-1}$ (Fig.~\ref{app21}(b)) and retreated to its most inland position (Fig.~\ref{app21}(c)).  In September 2005, the glacier has started to  decelerate with decreasing thinning rates (Fig.~\ref{app21}(a-b), while the terminus started re-advance (Fig.~\ref{app21}(c)) and by December the dynamic thinning rates dropped back to the 2001-2004 trend (Fig.~\ref{app21}(a), yellow band).

\subsubsection{ 2006-2010: pulsing behavior with intermittent thinning/thickening}
Following its readvance in the Fall of 2005, Helheim Glacier  exhibited an alternating thinning/thickening, retreat/advance, speed-up/slow-down behavior in 2006-2010 (Fig.~\ref{app21}(a-c)). Unlike in 2001-2004, seasonal variations featuring a simultaneous summer speed-up and retreat only occured in 2007 and 2010. 

\begin{figure}[h]
\centering
\includegraphics[width=3.0in]{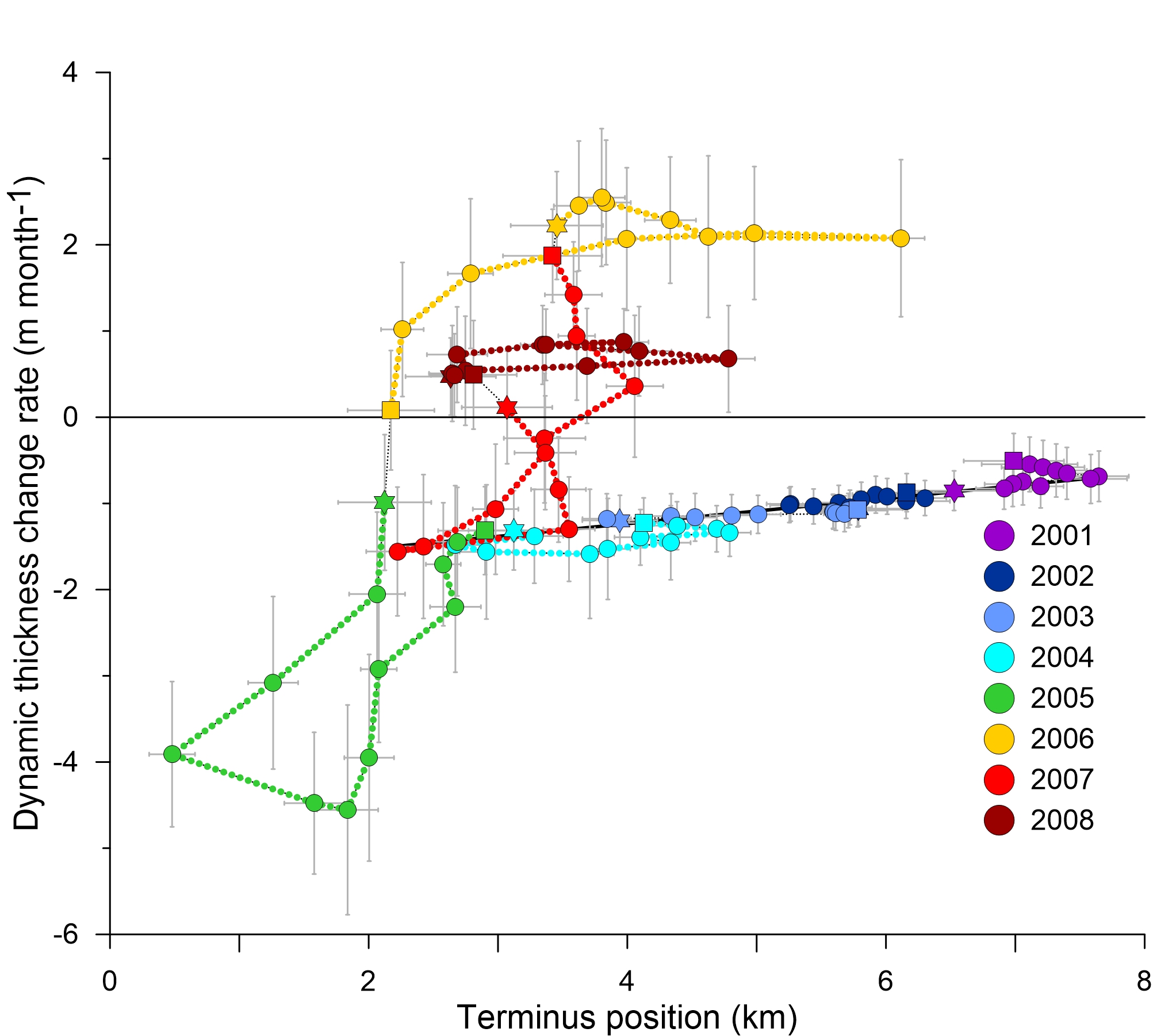}
\caption{Relationship between terminus position and the derivative of dynamic thickness  (dynamic thickness change rate) at TS:6a  between 2001 January 1- 2008 December 31 produced by ALPS. The values are shown for the first day of each calendar month with colors indicating the year, squares mark January and stars mark December of each year.  Error bars are according to 95 $\%$ CI of derivative determined by ALPS (grey band in Fig.~\ref{app21}(a)). Black line is linear approximation of trend,  selected months from 2001-2007.}
\label{app22}
\end{figure}

To investigate the main processes controlling the glacier behavior  during the different stages, we examined the relationship between the different parameters  and derivatives further.  Fig.~\ref{app22} shows the dynamic thickness change rates plotted against the terminus position  with different colors marking different years. We used the ALPS approximations of   dynamic thickness change rates    (Fig.~\ref{app21}(a), blue lines) and    terminus positions (Fig.~\ref{app21}(c), red lines) to calculate the monthly estimates shown in Fig.~\ref{app22}. 

The plot shows a clear relationship  between the terminus position and dynamic thickness change rates indicating that our analysis captured the behavior of the parameters at a suitable spatiotemporal scale. The following interpretation is based on the ALPS modeling result only.

In 2001-2004, thinning rates increased linearly as Helheim Glacier retreated inland (Fig.~\ref{app22}), black trendline). This behavior is consistent with  model results of Helheim's dynamic response to  the loss of buttressing, for example due to a positive feedback between increased calving and retreat  \cite{Nick:2014jp}. According to the least-squares fit of the linear trend,   dynamic thinning rates at TS:6c increased by -0.16 m month$^{-1}$ for every 1 km terminus retreat. 

The relationship between dynamic thinning/thickening and terminus retreat/advance altered dramatically  in April 2005. Between April and July 2005,  during a period of  small terminus variations,  glacier thinning increased very rapidly  (-2.1 m month$^{-1}$  for each 1 km retreat).  

A decrease of dynamic thinning in August 2005-January 2006 was followed by dynamic thickening as the glacier rapidly readvanced, almost reaching its 2001 terminus position by June 2006. 
During this period dynamic thickness change rates were mostly insensitive to the location of the terminus, perhaps because the glacier developed a floating tongue as it was suggested by  \cite{Joughin:2008cq}.   However, in 2007, the terminus retreat-dynamic thinning acceleration relationship  followed  the 2001-04 trend and  exhibited simultaneous summer speed-up and retreat (Figs.~\ref{app21}(a),~\ref{app22}),  suggesting a return to the 2001-04 behavior, perhaps due to the loss of the floating tongue.

Although glacier conditions (e.g., ice softening due to increased melt water) or bed geometry (retreat into deeper trough) can't be completely ruled out as the cause of the observed changes, we interpret the  rapid variations as the result of  change in the subglacial environment, either a sudden drop of  effective pressure, and thus a decrease in basal drag (e.g., \cite{Benn:2007hna}), or a weakening/strengthening of subglacial sediments under the glacier (e.g., \cite{Kulessa:2017bg}).

\section{ Conclusion} \label{conclusion}
This paper presents a novel modeling procedure for making inferences from time series data of land ice changes. The model is based on  penalized splines that are robust to errors and noise in the data and show much small bias in the fits. We provide a formal motivation for our approach by pointing out the disadvantages and issues with usage of global bases (like polynomials) and how our spline based approximation procedure can efficiently address all these issues and also provide uncertainty estimates for the predictions and estimates of rates of change. While explaining the details of our approach, we then provide a detailed reasoning of the decisions regarding the hyperparameters (like degree of basis functions and order of penalty) and how these decisions lead to a different modeling behavior on the chosen test sites from Greenland Ice Sheet. Finally, we show the successful application of our approach for increasing the resolution of the elevation change data by incorporating additional information from a physics based regional climate model that provide high-temporal resolution (10-day) estimates of ice sheet thickness changes caused by surface processes. We then use ALPS to model time series of dynamic ice thickness, ice velocity and terminus locations for  Helheim Glacier, a fast-flowing tidewater glacier. This example demonstrates the application of ALPS to  combine multisensor remote sensing data sets, obtained with different temporal sampling and characterized by different measurements errors, to investigate the relationship between the different expressions of outlet glacier dynamic behavior.  

In the future, we plan to extend ALPS on several aspects. Firstly, the uncertainty quantification can be made more accurate by using sampling based Bayesian approaches that are known to be more accurate \cite{ruppert2003semiparametric} (instead of the GCV approach implemented here). Then, we can also extend our approach to higher dimensions to model the behavior in full spatio-temporal approximation domain instead of just modeling time series at discrete locations. Finally, we also plan to extend the smoothing procedure by assuming a field for the hyperparameter ($\lambda$) and then making inference. This would allow us to incorporate other information like accuracy of individual sensors and in handling the sampling bias.

\section*{Acknowledgment}

The authors would like to thank Christopher Nuth for processing the MMASTER DEMs for Helheim Glacier, and Kristin Schild for sharing the 2001-2010 normalized terminus position time series of Helheim Glacier (pers. comm., 2018). The study was supported by NASA's Operation IceBridge and Sea Level Change science team grants, 
 NNX17AI65G and 80NSSC17K0611, respectively.

\ifCLASSOPTIONcaptionsoff
  \newpage
\fi



%

\bibliographystyle{IEEEtran}
\bibliography{IEEE_paper}

%
\vspace{-1cm}
\begin{IEEEbiography}
[{\includegraphics[width=1in,height=1.25in,clip,keepaspectratio]{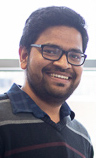}}]{Prashant Shekhar}
received the Bachelors degree in Industrial Engineering from Indian Institute of Technology, Kharagpur in 2014, the Masters degree in Mechanical Engineering from University at Buffalo in 2016 and the PhD. degree in Computational and Data Enabled Sciences also from University at Buffalo in 2019.
Currently his research focusses on learning hierarchical sparse representations for large datasets aimed at data reduction and efficient learning. He works with modeling datasets from cryosphere, remote sensing, numerical models and other related domains making joint inferences based on data and physics of the systems.
\end{IEEEbiography}
\vspace{-3cm}
\begin{IEEEbiography}
[{\includegraphics[width=1in,height=1.25in,clip,keepaspectratio]{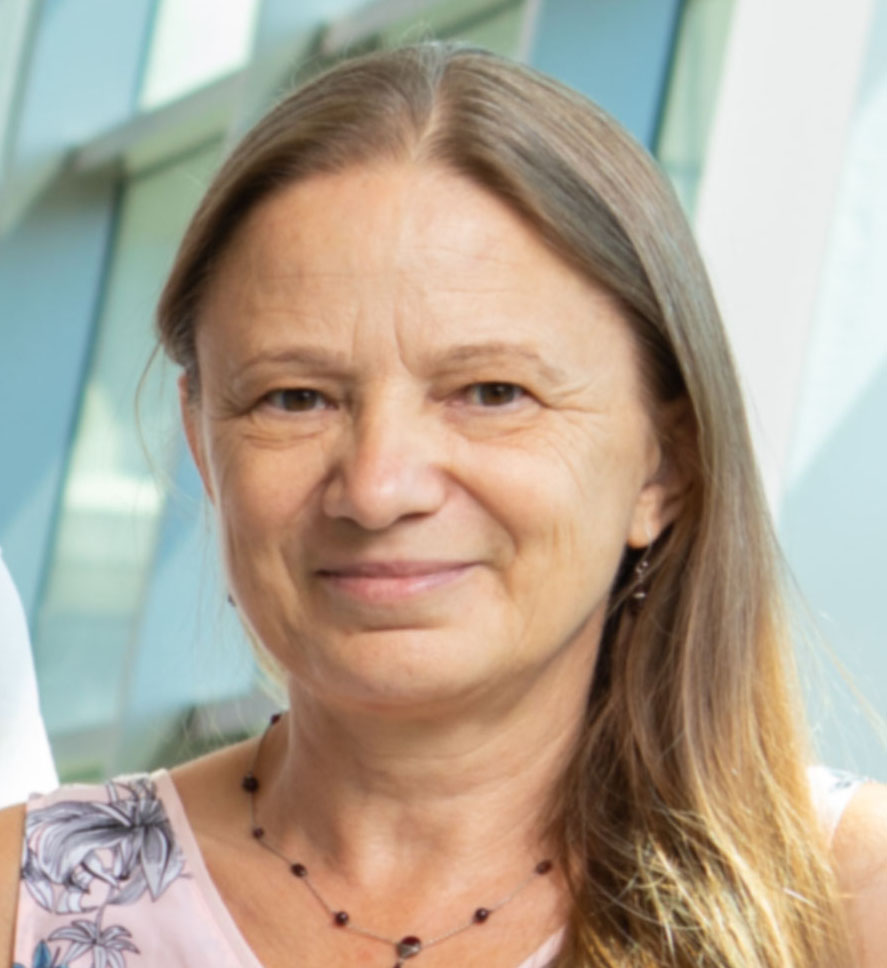}}]{Be\'ata Csath\'o}
Bea\'ata Csath\'o (M’00) received the M.S. degree in mathematics from E\"otv\"os Lor\'and University, Budapest, Hungary, in 1989 and the M.S. and Ph.D. degrees in geophysics from the University of Miskolc, Miskolc, Hungary, in 1981 and 1993. 
From 1981 to 1994, she was a Research Scientist with the E\"otv\"os Lor\'and Geophysical Institute, Budapest, Hungary and from 1994 to 2006 with the Byrd Polar Research Center, The Ohio State University, Columbus. In 2006, she joined the faculty of the Department of Geology, University at Buffalo, The State University of New York, Buffalo, where she is currently a Chair and Professor. She served as a Science Team Member of NASA's ICESat and ICESat-2 missions, and she is currently a science team member of the IceBridge mission. Her research interest  includes glaciology, remote sensing, geophysics,  spatial statistics,     and data fusion. 

Dr. Csath\'o is a member of the American Geophysical Union and the International Glaciological Society; She is a scientific editor of the Journal of Glaciology and a co-chair of ISPRS WG-III/9: Cryosphere and Hydrosphere. She was the recipient of a Fulbright Fellowship in 1992, the President\'s Honorary Citation and a Certificate of Appreciation from ISPRS in 2000 and 2001, and the NASA's Group Achievement Awards in 2004 and 2011. 
\end{IEEEbiography}
\vspace{-1cm}
\begin{IEEEbiography}
[{\includegraphics[width=1in,height=1.25in,clip,keepaspectratio]{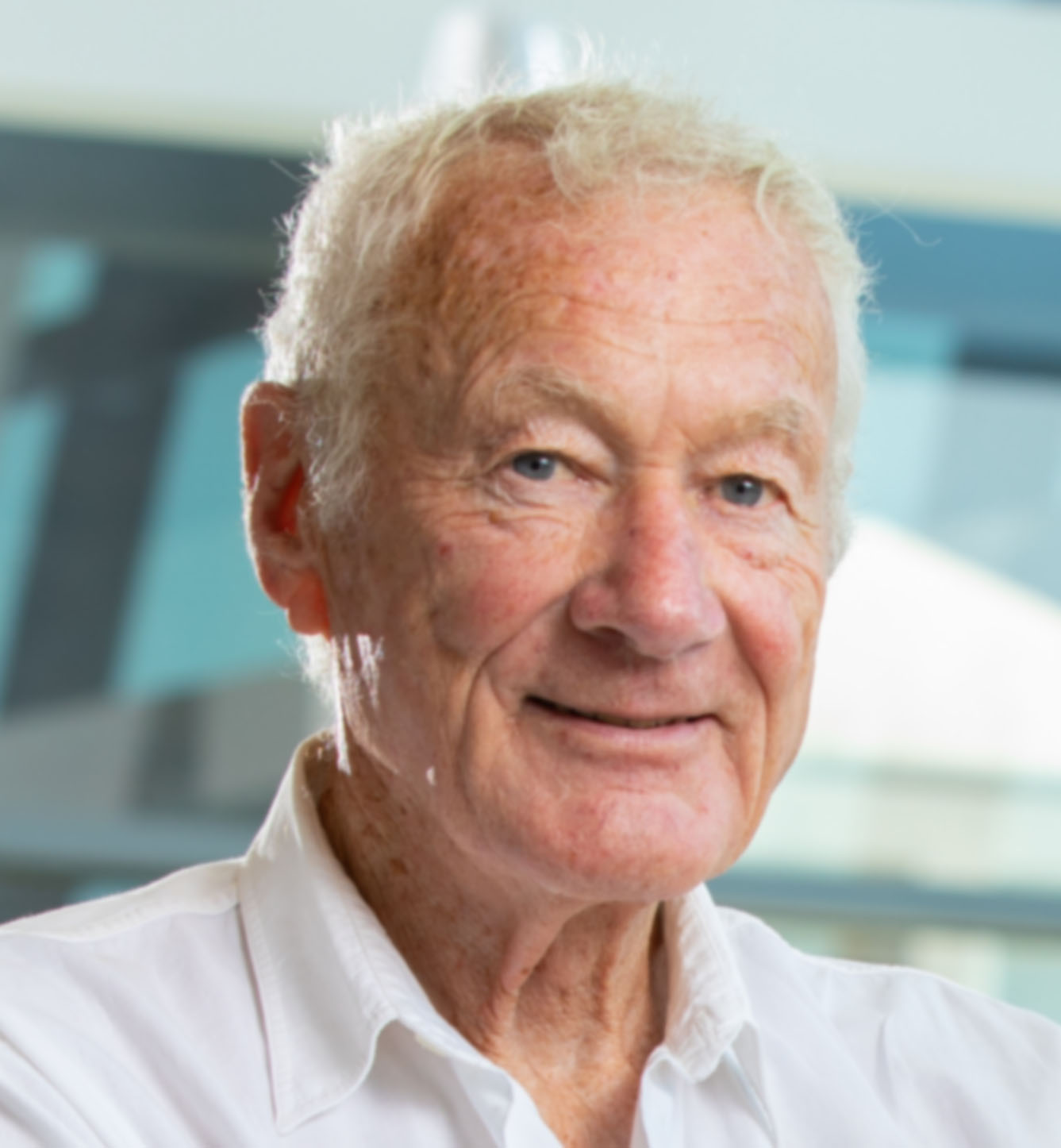}}]{Tony Schenk}
Tony Schenk received the M.S. degree in geodesy, photogrammetry, and cartography and the Ph.D. degree, both from the Swiss Federal Institute of Technology (ETH), Z\"urich, Switzerland, in 1965 and 1972, respectively. 
From 1972-1974, he was a Research Scientist at ETHZ before moving to California where he assumed the position of Operations Manager at California Aero Topo, Burlingame. Back in Switzeland, he became Division Head and Senior Scientist at Leica, Heerbrugg. In 1985, he accepted the position of Professor of photogrammetry at The Ohio State University, Columbus, where he worked until his retirement in 2010. He holds a Research Professor appointment at the Department of Geology, University at Buffalo, The State University of New York, Buffalo. He is author, coauthor, and editor of several books and has over 180 publications. His research interests are in digital photogrammetry, computer vision, and geospatial information science with an emphasis on object recognition and surface reconstruction. 

Dr. Schenk is a member of the American Society of Photogrammetry and Remote Sensing and the American Geophysical Union. He has hold numerous offices, including in the International Society of Photogrammetry and Remote Sensing, where he was President of a Technical Commission from 1996 to 2000. 
\end{IEEEbiography}
\begin{IEEEbiography}
[{\includegraphics[width=1in,height=1.25in,clip,keepaspectratio]{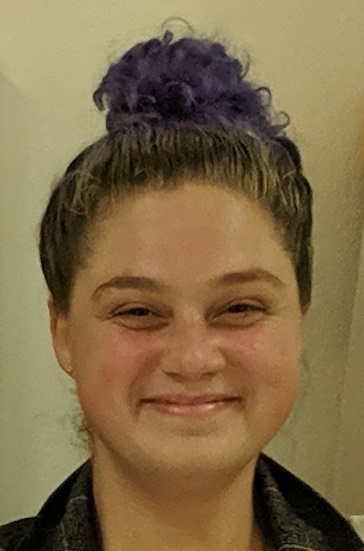}}]{Carolyn Roberts}
Carolyn Roberts employs remote sensing techniques to study planetary surface processes. She received her M.S. degree in Geological Sciences from the University at Buffalo in 2014, where she is currently a Research Assistant and has recently-defended her Ph.D.. Her previous projects include mapping lunar sinuous rilles, and lunar polar volatiles. She studies the ice surface in Greenland using laser altimetry and Digital Elevation Models, and performs science outreach in her spare time.
\end{IEEEbiography}
\begin{IEEEbiography}
[{\includegraphics[width=1in,height=1.25in,clip,keepaspectratio]{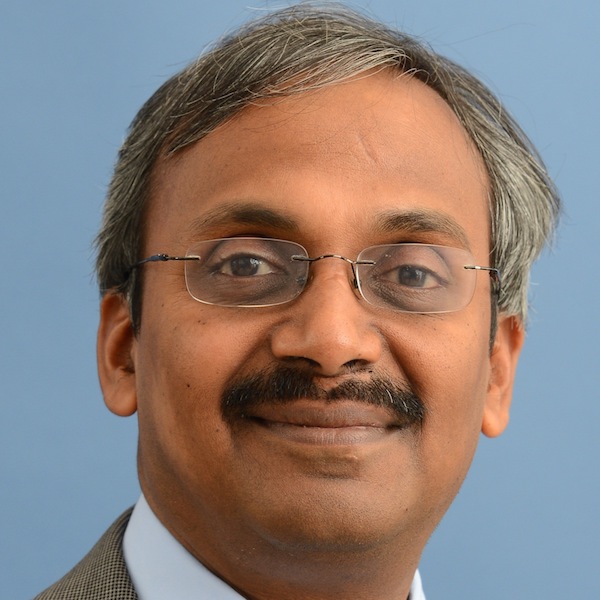}}]{Abani K Patra}
Abani K Patra is a computational and data scientist with a PhD in Computational and Applied Mathematics in 1995. He currently directs the multidisciplinary Data Intensive Studies Center at Tufts University in Medford, MA. His research interests include large scale computing, unertainty quantification, numerical analysis and mathematical modeling.
\end{IEEEbiography}






\end{document}